\documentclass[twocolumn,showpacs,preprintnumbers,superscriptaddress,amsfonts,prd,floatfix,tightenlines]{revtex4}
\usepackage{graphicx}
\usepackage{dcolumn}
\usepackage{longtable}%
\usepackage[dvips]{epsfig}
\usepackage{bm}
\newcommand{\piz}{\mbox{$\pi^{0}$}}
\newcommand{\pim}{\mbox{$\pi^{-}$}}
\newcommand{\kmi}{\mbox{$K^{-}$}}
\newcommand{\ycm}{\mbox{$y_{\rm{cm}}$}}
\def\PRETH{{\sc pretrigger hi}}
\def\GLL{{\sc local$\otimes$global lo}}

\def\SLL{{\sc single local lo}}
\def\SLH{{\sc single local hi}}
\def\INT{{\sc interaction}}
\def\BEAM{{\sc beam}}
\def\BEAMONE{{\sc beam1}}
\def\DIFFXS{Ed \sigma/d^{3}p}

\def\CHERENKOV{Cherenkov}
\def\LOCALH{{\sc local hi}}
\def\LOCALL{{\sc local lo}}

\def\GLOBALL{{\sc global lo}}

\def\avkt{$\langle k_T \rangle$}
\def\pt{\mbox{$p_T$}}
\def\xt{\mbox{$x_T$}}
\def\kt{$k_T$}
\def\pBe{${\mit p}$Be}
\def\pp{${\mit pp}$}
\def\pip{$\pi^{-}\mit{p}$}
\def\piBe{${\pi^{-}}$Be}

\def\vctr#1{\vbox to-2ex{\vss\hbox{#1}\vss}}%
\def\UCDAVIS{University of California-Davis, Davis, California 95616}
\def\MSU{Michigan State University, East Lansing, Michigan 48824}
\def\Delhi{University of Delhi, Delhi, India 110007}
\def\FNAL{Fermi National Accelerator Laboratory, Batavia,
                   Illinois 60510}
\def\NEU{Northeastern University, Boston, Massachusetts  02115}
\def\OK{University of Oklahoma, Norman, Oklahoma  73019}
\def\PSU{Pennsylvania State University, University Park,
		   Pennsylvania 16802}
\def\PU{University of Pittsburgh, Pittsburgh, Pennsylvania 15260}
\def\UR{University of Rochester, Rochester, New York 14627}
\def\bbcoord{20 150 652 652} 
\def\bbcoorda{20 150 652 652} 
\def\bbcoordb{50 150 652 652} 
\def\bbcoordc{50 220 652 740} 
\def\bbcoordd{50 400 652 675}
\def\bbcoorde{20 100 652 720}  
\def\figsize{4.0in}

\begin{document}

\preprint{FERMILAB-Pub-02/230-E}
\title{
Production of $\bm \pi^0$ and $\bm \eta$ mesons at large transverse momenta \\
in $\bm \pi^{-}\mit{p}$ and $\bm \pi^-$Be interactions at 515~GeV/$c$}
\author{L.~Apanasevich}\affiliation{\MSU}\affiliation{\UR}
\author{J.~Bacigalupi}\affiliation{\UCDAVIS}
\author{W.~Baker}\affiliation{\FNAL}
\author{M.~Begel}\affiliation{\UR}
\author{S.~Blusk}\affiliation{\PU}
\author{C.~Bromberg}\affiliation{\MSU}
\author{P.~Chang}\affiliation{\NEU}
\author{B.~Choudhary}\affiliation{\Delhi}
\author{W.~H.~Chung}\affiliation{\PU}
\author{L.~de~Barbaro}\affiliation{\UR}
\author{W.~DeSoi}\affiliation{\UR}
\author{W.~D{\l}ugosz}\affiliation{\NEU}
\author{J.~Dunlea}\affiliation{\UR}
\author{E.~Engels,~Jr.}\affiliation{\PU}
\author{G.~Fanourakis}\affiliation{\UR}
\author{T.~Ferbel}\affiliation{\UR}
\author{J.~Ftacnik}\affiliation{\UR}
\author{D.~Garelick}\affiliation{\NEU}
\author{G.~Ginther}\affiliation{\UR}
\author{M.~Glaubman}\affiliation{\NEU}
\author{P.~Gutierrez}\affiliation{\OK}
\author{K.~Hartman}\affiliation{\PSU}
\author{J.~Huston}\affiliation{\MSU}
\author{C.~Johnstone}\affiliation{\FNAL}
\author{V.~Kapoor}\affiliation{\Delhi}
\author{J.~Kuehler}\affiliation{\OK}
\author{C.~Lirakis}\affiliation{\NEU}
\author{F.~Lobkowicz}\altaffiliation{Deceased}\affiliation{\UR}
\author{P.~Lukens}\affiliation{\FNAL}
\author{S.~Mani~Tripathi}\affiliation{\UCDAVIS}
\author{J.~Mansour}\affiliation{\UR}
\author{A.~Maul}\affiliation{\MSU}
\author{R.~Miller}\affiliation{\MSU}
\author{B.~Y.~Oh}\affiliation{\PSU}
\author{G.~Osborne}\affiliation{\UR}
\author{D.~Pellett}\affiliation{\UCDAVIS}
\author{E.~Prebys}\affiliation{\UR}
\author{R.~Roser}\affiliation{\UR}
\author{P.~Shepard}\affiliation{\PU}
\author{R.~Shivpuri}\affiliation{\Delhi}
\author{D.~Skow}\affiliation{\FNAL}
\author{P.~Slattery}\affiliation{\UR}
\author{L.~Sorrell}\affiliation{\MSU}
\author{D.~Striley}\affiliation{\NEU}
\author{W.~Toothacker}\altaffiliation{Deceased}\affiliation{\PSU}
\author{N.~Varelas}\affiliation{\UR}
\author{D.~Weerasundara}\affiliation{\PU}
\author{J.~J.~Whitmore}\affiliation{\PSU}
\author{T.~Yasuda}\affiliation{\NEU}
\author{C.~Yosef}\affiliation{\MSU}
\author{M.~Zieli\'{n}ski}\affiliation{\UR}
\author{V.~Zutshi}\affiliation{\Delhi}
\collaboration{Fermilab E706 Collaboration}\noaffiliation

\date{\today}

\begin{abstract}
We present results on the production of high transverse momentum
$\pi^0$ and $\eta$ mesons in \pip\ and $\pi^-$Be interactions at
515~GeV/$c$. The data span the kinematic ranges $1<p_T<11$~GeV/$c$ in
transverse momentum and $-0.75 \le \ycm\ \le 0.75$ in rapidity.  The
inclusive $\pi^0$ cross sections are compared with next-to-leading
order QCD calculations and to expectations based on a phenomenological
parton-$k_T$ model.
\end{abstract}

\maketitle

\section{Introduction}

The study of inclusive single-hadron production at large transverse
momentum ($p_T$) has been a useful probe in the development of
perturbative QCD (PQCD)~\cite{geist,mccubbin}.  Early in the evolution
of the parton model, a departure from an exponential dependence of
particle production at low $p_T$ was interpreted in terms of the onset
of interactions between pointlike constituents (partons) contained in
hadrons.  Large $p_T$ is a regime where perturbative methods have been
applied to QCD to provide quantitative comparisons with data.  Such
comparisons yield information on the validity of the PQCD description,
and on the parton distribution functions of hadrons and the
fragmentation functions of partons.

This paper reports high-precision measurements of the production of
$\pi^0$ and $\eta$ mesons with large $p_T$ by 515~GeV/$c$ \pim\ beams.
The data were accumulated during the 1990 and 1991-92 fixed-target runs
at Fermilab.  The $\pi^0$ production cross sections are compared with
next-to-leading order (NLO) PQCD calculations~\cite{aversa}.  As
illustrated in a previous publication~\cite{prl}, our data for both
inclusive $\pi^0$ and direct-photon production are not described
satisfactorily by the available NLO PQCD calculations, using standard
choices of parameters.  Similar discrepancies have been
observed~\cite{prd} between conventional PQCD calculations and other
measurements of high-$p_T$ $\pi^0$ and direct-photon cross sections
(see also \cite{huston,patrick,frenchpiz}).  The origin of these
discrepancies can be attributed to the effects of initial-state
soft-gluon radiation.  Such radiation generates transverse components
of initial-state parton momenta, referred to below as
$k_T$~\cite{ktnote}.  Evidence of significant $k_T$ in various
processes, and a phenomenological model for incorporating its effect
on the calculated high-$p_T$ cross sections, have been extensively
discussed in Refs.~\cite{prl,prd}. Recent studies of the
photoproduction of direct photons at HERA may provide additional
insights~\cite{ZEUS-dp,ZEUS-kt,fontannaz-c21,fontannaz-c22,zembrzuski}.
In this paper, we follow the phenomenological prescription of 
Ref.~\cite{prd} when comparing calculations with our \piz\ data.

\section{Experimental Apparatus}

\subsection{Spectrometer}

\begin{figure}
\epsfxsize=\figsize \epsfbox[\bbcoordc]{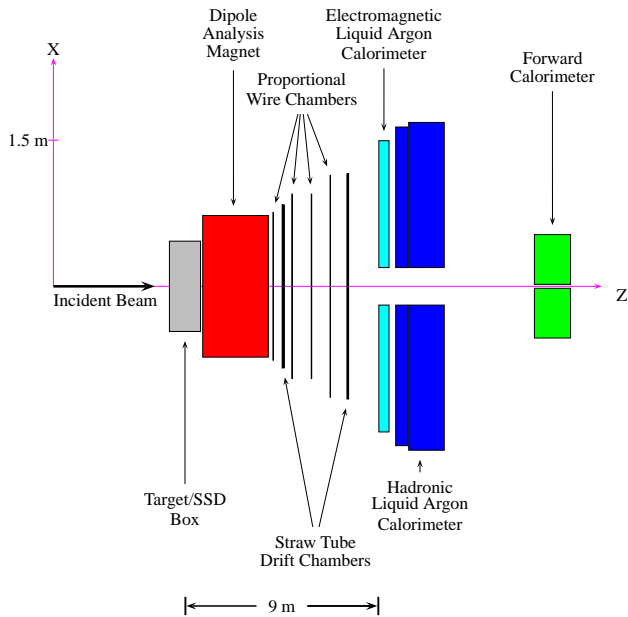}
\caption{Plan view of the Meson West spectrometer. The $Z$-axis of the
coordinate system for the experiment was oriented along the nominal beam
axis.}
\label{fig:e706-spect}
\end{figure}

Fermilab E706 was a fixed-target experiment designed to measure the
production of direct photons, neutral mesons, and associated particles
at high-$p_T$~\cite{prl,E706-pi-eta,E706-omega,E706-charm}. The
spectrometer, designed and built for this experiment, was located in
the Meson West experimental hall and included a precision charged
particle tracking system and a large acceptance liquid argon
calorimeter. Figure~\ref{fig:e706-spect} shows the key elements of the
Meson West spectrometer~\cite{E672-spec}.

The tracking system consisted of silicon microstrip detectors (SSDs)
in the target region, and multiwire proportional chambers (PWCs) and
straw tube drift chambers (STDCs) downstream of a large aperture
analysis magnet~\cite{E706-charm}.  The SSD system contained sixteen
planes of silicon wafers, arranged in eight modules. Each module
contained two SSD planes, one providing $X$-view information and the
other providing $Y$-view information.  Six 3$\times$3~cm$^2$ SSD
planes were located upstream of the target and used to reconstruct
beam tracks.  Two hybrid 5$\times $5~cm$^2$ SSD planes (25~$\mu$m
pitch strips in the central 1~cm; 50~$\mu$m beyond) were located
downstream of the target. These were followed by eight 5$\times
$5~cm$^2$ SSD planes of 50~$\mu$m pitch.  The analysis dipole imparted
a transverse momentum impulse in the horizontal plane of
$\approx 450~{\rm MeV}/c$ to charged particles.  Downstream track
segments were measured by means of four stations of four views
($XYUV$) of 2.54~mm pitch PWCs and two stations of eight (4$X$4$Y$)
layers of STDCs with tube diameters 1.04~cm (upstream station) and
1.63~cm (downstream station).

\begin{figure}
\epsfxsize=\figsize \epsfbox[\bbcoorda]{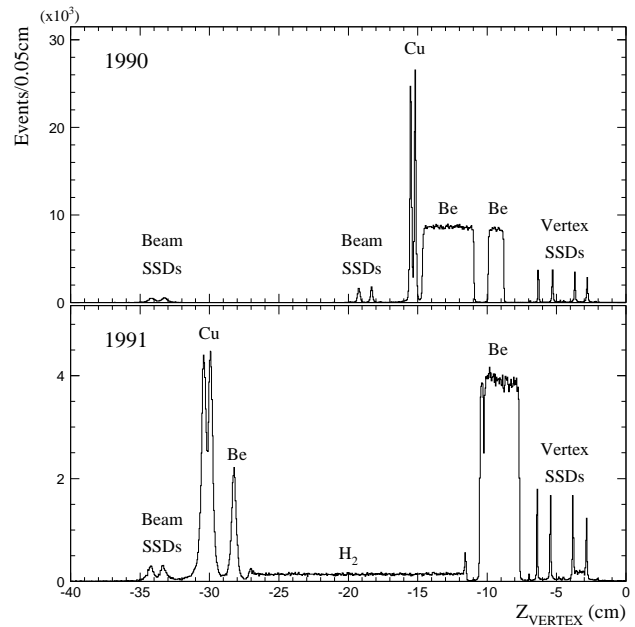}
\caption{The longitudinal distribution of reconstructed vertices for the
1990~(top) and 1991-92~(bottom) target configurations.}
\label{fig:vz}
\end{figure}

In the 1990 run, the target consisted of two 0.8~mm thick copper foils
followed by two pieces of beryllium. The upstream Be piece was 3.7~cm
long, while the downstream Be piece was 1.1~cm long. For the 1991-92
run, the target was reconfigured to include a liquid hydrogen
target~\cite{h2target} contained in a 15.3~cm long mylar flask and
supported in an evacuated volume with beryllium windows at each end
(2.5~mm thickness upstream and 2.8~mm thickness downstream).  The
liquid hydrogen target was flanked by two 0.8~mm thick copper disks
upstream, and a 2.54~cm long beryllium cylinder downstream.
Figure~\ref{fig:vz} shows the reconstructed vertex position along $Z$
as determined by the tracking system for representative samples from
the 1990 (top) and 1991-92 (bottom) runs.  The individual target
elements are clearly resolved, as are several of the SSDs.

Photons were detected in a large, lead and liquid-argon sampling
electromagnetic calorimeter (EMLAC), located 9~m downstream of the
target~\cite{E706-calibration}.  The EMLAC had a cylindrical geometry
with an inner radius of 20~cm and an outer radius of 160~cm.  It was
divided into four mechanically independent quadrants, which were
further subdivided electronically to create octants.  The calorimeter
had 33~longitudinal cells, read out in two sections: an 11 cell front
section ($\approx 8.5$~radiation lengths) and a 22 cell back section
($\approx 18$~radiation lengths).  The longitudinal cells consisted of
2~mm thick lead cathodes (the first cathode was constructed of
aluminum), double-sided copper-clad G-10 radial ($R$) anode boards
followed by 2~mm thick lead cathodes, and double-sided copper-clad
G-10 azimuthal ($\Phi$) anode boards. There were 2.5~mm argon gaps
between each of these layers in a cell. The copper-cladding on the
anode boards was cut to form strips. Signals from corresponding strips
on all $R$ (or $\Phi$) anode boards in the front (or back) section
were jumpered together.  The copper-cladding on the radial anode
boards was cut into concentric strips centered on the nominal beam
axis. The width of the strips on the first $R$ board was 5.5~mm. The
width of the $R$ strips on the following $R$ boards increased slightly
so that the radial geometry was projective relative to the target. The
azimuthal readout was subdivided at a radius of 40~cm into inner and
outer segments, with each inner $\Phi$ strip subtending an azimuthal
angle of $\pi/192$~radians, and outer $\Phi$ strips covering
$\pi/384$~radians.  Subdivision of the azimuthal strips in the outer
portion of the detector improved both the position and energy
resolution for showers reconstructed in this region.  It also reduced
$R$--$\Phi$ correlation ambiguities from multiple showers in the same
octant of the calorimeter.

The spectrometer also included two other calorimeters: a hadronic
calorimeter located downstream of the EMLAC within the same cryostat,
and a steel and scintillator calorimeter positioned further downstream
to provide instrumentation in the very forward region.  The E672 muon
spectrometer, consisting of a toroidal magnet, scintillators, and
proportional wire chambers, was deployed immediately downstream of the
calorimeters~\cite{B90,psi90,chi90,psi91}.

The beamline was instrumented with a differential
\CHERENKOV\ counter~\cite{striley,kourbanis} to identify incident
pions, kaons, and protons in secondary beams.  This helium-filled
counter was 43.4~m long and was located $\approx 100$~m upstream of the
experiment's target. At the end of the beamline was a 4.7~m long stack
of steel surrounding the beam pipe and shadowing the EMLAC to absorb
off-axis hadrons. A water tank was placed at the downstream end of
this hadron shield to absorb low-energy neutrons. Surrounding the
hadron shield and neutron absorber were walls of scintillation
counters (VW) to identify penetrating muons.  During the 1990 run,
there was one wall at the upstream end and two walls at the downstream
end of the hadron absorber. During the 1991-92 run, an additional wall
was added to the upstream end.

\subsection{Trigger}

The E706 trigger selected events yielding high transverse momentum
showers in the EMLAC. The selection process involved several stages:
beam and interaction definitions and the pretrigger and high-$p_T$
trigger requirements. Beam particles were detected using a hodoscope
consisting of three planes (arranged in $X$, $Y$, and $U$ views) of
scintillator strips, located $\approx 2$~m upstream of the target
region. A \BEAM\ signal was generated if the beam hodoscope registered
hits in time coincidence from at least two of the three planes. In
addition, a \BEAMONE\ signal was generated if the hits in at least two
of the hodoscope planes were considered to be isolated.  Scintillation
counters centered on the target with a 0.95~cm diameter central hole
were used to reject interactions initiated by particles in the beam
halo.

Interactions were detected using two pairs of scintillation counters
mounted on the dipole analysis magnet, one pair on the upstream side
and one pair downstream.  Each pair had a central hole that allowed
non-interacting beam particles to pass through undetected. An \INT\
was defined as a coincidence between signals from at least two of
these four interaction counters. A filter was used to reject
interactions that occurred within 60~ns of one another to minimize
potential confusion in the EMLAC due to out-of-time interactions.

For those interactions that satisfied both the \BEAMONE\ and \INT\
definitions, the $p_T$ deposited in various regions of the EMLAC was
evaluated by weighting the energy signals from the fast outputs of the
EMLAC $R$ channel amplifiers by $\approx \sin{\theta_i}$, where
$\theta_i$ is the polar angle that the $i^{th}$ strip subtended with
respect to the nominal beam axis.  The \PRETH\ signal required that
the $p_T$ detected in either the inner or the outer $R$ channels of
any octant was greater than the pretrigger threshold value. This
signal was issued only when there was no evidence in that octant of
substantial noise or significant \pt\ attributable to an earlier
interaction, and that there was no incident beam halo muon detected by
the VW.

Localized trigger groups were formed for each octant by clustering the
inputs from the $R$ channels into 32~groups of 8~channels.  Each of
the adjacent pairs of 8~channel groups formed a group of 16 strips.
If the $p_T$ detected in any of these groups of 16 was above a
specified high (or low) threshold, then a \LOCALH\ (or \LOCALL) signal
was generated for that octant. If a \LOCALH\ (or \LOCALL) signal was
generated in coincidence with the \PRETH\ in the same octant, then a
\SLH\ (or \SLL) trigger was generated for that octant. 

Trigger decisions were also made based upon global energy
depositions in an octant. A \GLOBALL\ signal was generated if 
the total \pt\ in an octant exceeded a low global threshold 
value. The \GLL\ trigger required a coincidence of the
\GLOBALL\ signal with \PRETH\ and \LOCALL\ signals from the 
same octant. The \LOCALL\ requirement was included to suppress
spurious global triggers due to coherent noise in the EMLAC.

The \SLL\ and \GLL\ triggers were prescaled to keep them from
dominating the data sample. Prescaled samples of beam, interaction,
and pretrigger events were also recorded.  Further details concerning
the E706 trigger can be found
elsewhere~\cite{E706-pi-eta,E706-charm,E706-trigger,osborne}.

\section{Analysis Methods}

This paper presents results for \piz\ and $\eta$ production by
515~GeV/$c$ $\pi^-$ beam on beryllium and liquid hydrogen targets.
The data sample used in this analysis corresponds to an integrated
luminosity of 7.5~$pb^{-1}$ of $\pi^-$Be data (6.1~$pb^{-1}$ from the
1990 run plus 1.4~$pb^{-1}$ from the 1991-92 run) and 0.23~$pb^{-1}$
of $\pi^-p$ data. The following subsections describe the data analysis
procedures and the methods used to correct the data for losses due to
inefficiencies and selection biases.  Additional details may be found
in several of our previous
papers~\cite{E706-pi-eta,E706-charm,E706-calibration}.

\subsection{Reconstruction}

Two major aspects of the event reconstruction procedure were particle
track and calorimeter shower reconstruction.  The charged-track
reconstruction algorithm produced track segments upstream of the
magnet using information from the SSDs, and downstream of the magnet
using information from the PWCs and STDCs. These track segments were
projected to the center of the magnet and linked to form the final
reconstructed tracks, whose calculated charges and momenta were used
in physics analyses and to determine the location of the primary
interaction vertex.  The charged track reconstruction is
described in detail in Ref.~\cite{E706-charm}.

The readout in each EMLAC quadrant consisted of four regions: left $R$
and right $R$, (radial strips of each octant in that quadrant), and
inner $\Phi$ and outer $\Phi$.  Strip energies from clusters in each
region were fit to the shape of an electromagnetic shower as
determined from Monte Carlo simulations and isolated-shower data.
These fits were used to evaluate the positions and energies of the
peaks in each region.  Shower positions and energies were obtained by
correlating peaks of approximately the same energy in the $R$ and
$\Phi$ regions within the same half octant (more complex algorithms
were used to handle configurations with overlapping showers in either
the $R$ or $\Phi$ regions).  The EMLAC readout was also subdivided
longitudinally into front and back sections.  This longitudinal
segmentation provided discrimination between showers generated by
electromagnetically or hadronically interacting particles.  An
expanded discussion of the EMLAC reconstruction procedures and
performance can be found in Ref.~\cite{E706-calibration}.

\subsection{Event selection and meson signals}

Events contributing to measurements of cross sections were required to
have reconstructed vertices within the fiducial volume of the Be or
H${}_2$ targets. Both $\pi^0$ and $\eta$ mesons were reconstructed via
their $\gamma\gamma$ decay modes.  Only those showers which deposited
at least $20\%$ of their energy in the front part of EMLAC were
considered as $\gamma$ candidates, to reduce the background due to
showers from hadronic interactions.  Only those $\gamma\gamma$
combinations with energy asymmetry $A_{\gamma\gamma} \equiv |E
_{\gamma_1}-E_{\gamma_2}|/ (E_{\gamma_1}+ E_{\gamma_2})<0.75$ were
considered in order to reduce uncertainties due to low energy photons.
Photons were required to be reconstructed within the fiducial region
of the EMLAC to exclude areas with reduced sensitivity.  In
particular, regions of the detector near quadrant boundaries (which
abutted steel support plates), the central beam hole, the outer radius
of the EMLAC, and octant boundaries were excluded.  In addition,
$\gamma\gamma$ combinations were restricted to the same octant to
simplify the trigger analysis.  A simple ray-tracing Monte Carlo
program was employed to determine the correction for these fiducial
requirements.

Signals have been corrected for the $25\%$ loss due to the energy asymmetry 
cut and the branching fractions for the $\gamma\gamma$ decay
modes~\cite{pdg}.  The correction for losses due to the conversion of
one or both of the photons into $e^+e^-$ pairs was evaluated by
projecting each reconstructed photon from the event vertex to the
reconstructed position in the EMLAC.  The radiation length of
material traversed was calculated based upon detailed descriptions
of the detectors encountered. The photon conversion probability was 
evaluated and used to account for conversion losses. The average 
correction for conversion losses was $1.19$ for the Be target in the 
1990 run ($1.16$ in 1991-92) and $1.24$ for the $p$ target. A full event 
simulation (described below) was employed to correct for other effects
including reconstruction smearing and losses.

\begin{figure}[t]
\epsfxsize=\figsize \epsfbox[\bbcoordd]{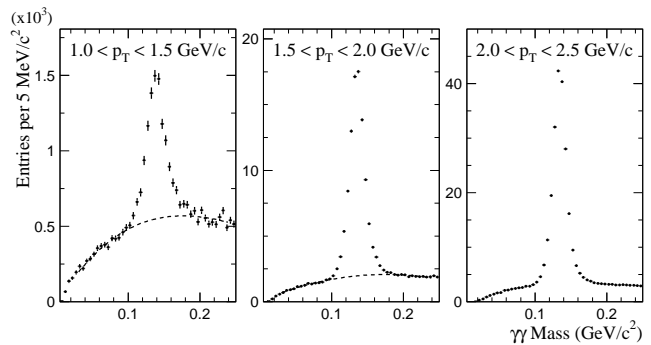}
\caption{$\gamma\gamma$ mass distributions in the region of the 
$\pi^0$ for several $p_T$ bins.
Curves are overlayed for those \pt\ bins where the background to the
signal was determined using a fitting procedure rather than sideband 
subtraction.}
\label{fig:mvpt_pi0}
\end{figure}

\begin{figure}
\epsfxsize=\figsize \epsfbox[\bbcoordd]{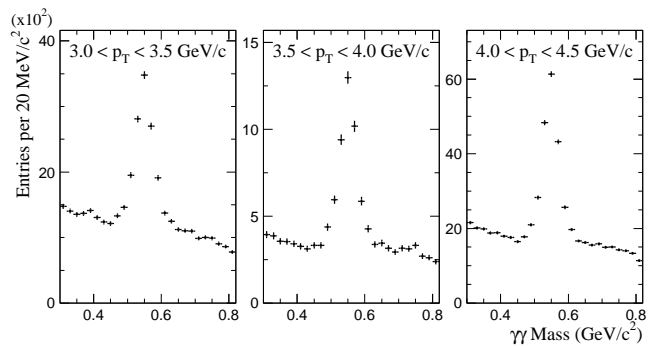}
\caption{$\gamma\gamma$ mass distributions in the region of the 
$\eta$ for several $p_T$ bins.}
\label{fig:mvpt_eta}
\end{figure}

The $\gamma\gamma$ invariant mass distributions in the $\pi^0$ and
$\eta$ mass regions for photon-pairs that satisfied the above
requirements are shown in Figs.~\ref{fig:mvpt_pi0} and
\ref{fig:mvpt_eta} for several $\gamma\gamma$ $p_T$ intervals.  A
$\pi^0$ candidate was defined as a combination of two photons with
invariant mass, $M_{\gamma\gamma}$, in the range $100~{\rm
MeV}/c^2<M_{\gamma\gamma}<180~{\rm MeV}/c^2$. An $\eta$ candidate was
defined as a two-photon combination in the range $450~{\rm
MeV}/c^2<M_{\gamma\gamma}<650~{\rm MeV}/c^2$.  Combinatorial
background under the $\pi^0$ and $\eta$ peak regions was evaluated as
follows. Sideband regions were defined to cover a mass range
equivalent to that in the $\pi^0$ and $\eta$ peak regions.
Distributions from these side bands were subtracted from the
distributions within the $\pi^0$ and $\eta$ mass ranges to obtain the
respective signals.  This technique is appropriate as long as the
combinatorial background depends approximately linearly on
$M_{\gamma\gamma}$.  The combinatorial background shape is not linear
at low $p_T$ (below $\approx 2$~GeV/$c$), and a fitting procedure was
used to evaluate this background. The $\gamma\gamma$ mass
distributions at low $p_T$ were fit using Gaussians for signal, and
second and third order polynomials for the background.  The
combinatorial background in the peak regions was determined from the
resultant fit parameters, and subsequently subtracted from the total
entries within the peak.

\subsection{Trigger Corrections}

Trigger corrections were evaluated on an event-by-event basis using
the measured efficiencies of the trigger groups responsible for the
formation of a given trigger. For example, the \SLH\ trigger
corrections were based upon the efficiencies of the 32
groups of 16 in the triggering octant. These efficiencies were
evaluated as functions of the \pt\ reconstructed within the trigger
group, using data samples that were unbiased with respect to the
trigger group.  From these efficiencies, a trigger probability was
defined, $P = 1 -\prod(1-p_{\mit i})$, where $p_{\mit i}$ is the
efficiency of the ${\mit i}^{th}$ trigger group in the octant. The
inverse of this probability was applied as a trigger weight to each
meson candidate. Meson candidates with trigger probabilities of $P <
0.1$ were excluded from further consideration to avoid excessively
large trigger weights. The correction for losses from this requirement
was determined from Monte Carlo, and absorbed into the reconstruction
efficiency.

The cross sections presented in this paper utilize the results from 
a combination of triggers. Going from low to high \pt, the 
triggers were: \INT, \PRETH, \GLL\ (1990 run) or \SLL\ 
(1991-92 run), and \SLH. The transition points chosen
between the lower and higher threshold triggers were determined by
comparing the fully corrected results from each trigger, and were
different for \piz\ and $\eta$ mesons, and also depended on rapidity.
For \pt$ > 4.0$ GeV/c, the \SLH\ trigger was used exclusively for both 
\piz\ and $\eta$ mesons.

\subsection{Beam halo muon rejection}

\begin{figure}
\epsfxsize=\figsize \epsfbox[\bbcoordb]{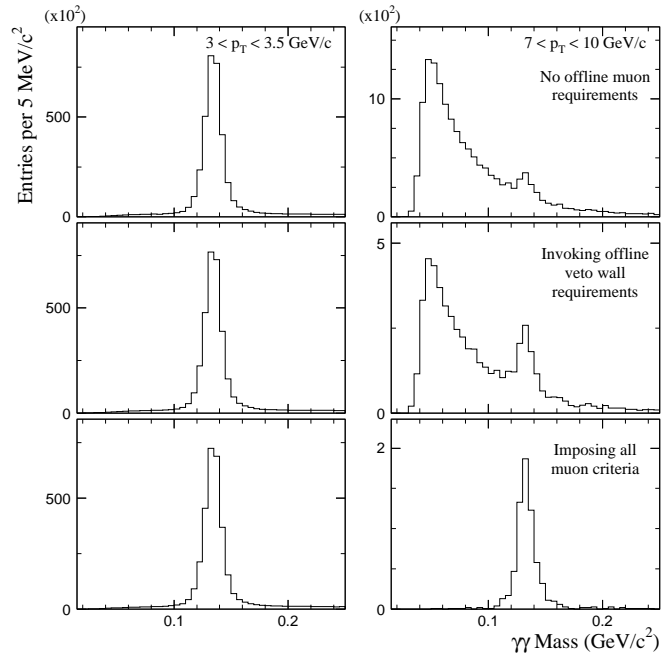}
\caption{Effect of muon rejection requirements on the invariant mass 
distribution in the \piz-mass region, for candidate $\gamma\gamma$  
pairs with $3 < p_T < 3.5$~GeV/$c$ (left) and $7 < p_T < 10$~GeV/$c$ (right).}
\label{fig:muons}
\end{figure}

Spurious triggers were produced by muons in the beam halo that
deposited energy in the electromagnetic calorimeter in random
coincidence with an interaction in the target. Particularly in the
outer regions of the EMLAC, such energy depositions can produce
background at low $\gamma\gamma$ mass values due to the occasional
splitting of the muon-induced showers into two closely-separated
photon candidates. The pretrigger logic used signals from the VW to
reject events associated with such muons in the beam halo. The
off-line analysis employed expanded requirements on the latched VW
signals, and imposed requirements upon the direction of reconstructed
showers, the shower shape, and the total $p_T$ imbalance in the event
to futher reduce this background.  The effects of these off-line
rejection requirements on the $\gamma\gamma$ invariant mass
distribution are shown in Fig.~\ref{fig:muons}, for $\gamma\gamma$
pairs with $3 < p_T < 3.5$~GeV/c (left) and $7 < p_T < 10$~GeV/c
(right). The rejection requirements completely eliminate the large
muon-induced background in the high-\pt\ bin while having very little
effect on the signal.  A more detailed description of these
requirements can be found in Ref.~\cite{E706-pi-eta}.

\subsection{Detector simulation}

The Meson West spectrometer was modeled via a detailed {\sc
geant}~\cite{geant} simulation.  A preprocessor was used to convert
{\sc geant} information into the hits and strip energies measured by
the various detectors.  The preprocessor simulated hardware effects,
such as channel noise and gain variations.  Monte Carlo generated
events were then processed through the same reconstruction software
used to analyze the data.  This accounted for inefficiencies
and biases in the reconstruction algorithms.  Reconstruction
inefficiencies for $\pi^0$ and $\eta$ mesons were relatively small
over most of the kinematic range.  More information on the detailed
simulation of the Meson West spectrometer can be found in
Ref.~\cite{E706-pi-eta}. We employed single particle distributions,
reconstructed data, and the {\sc herwig}~\cite{herwig56} physics
generator as inputs to the {\sc geant} simulations.  {\sc herwig}
calculations of $\pi^0$ and $\eta$ spectra were weighted in $p_T$ and
rapidity using our measured results in an iterative fashion so that
the final corrections were based on the data distributions rather than
on the output of the physics generator. 

Spectral effects were particularly important to the calibration of the
EMLAC's energy response~\cite{E706-calibration}.  The calibration of
the energy response was based on the reconstructed masses of $\pi^0$
mesons in the $\gamma\gamma$ decay mode.  The steeply falling $\pi^0$
$p_T$ spectrum, combined with the calorimeter's resolution, results in
a small offset ($\approx 1\%$) in the mean reconstructed photon
energies.  We accounted for this offset, and for potential biases in
the calibration procedure, by calibrating the simulated EMLAC in the
same manner as the real detector.  We also employed the simulation to
evaluate the mean correction (as a function of photon energy) for
energy deposited in the material upstream of the EMLAC.  The impacts
of detector resolution on the energy scale calibration and on the
$\pi^0$ and $\eta$ production spectra were incorporated in the overall
reconstruction efficiency corrections.

\begin{figure}
\epsfxsize=\figsize \epsfbox[\bbcoordb]{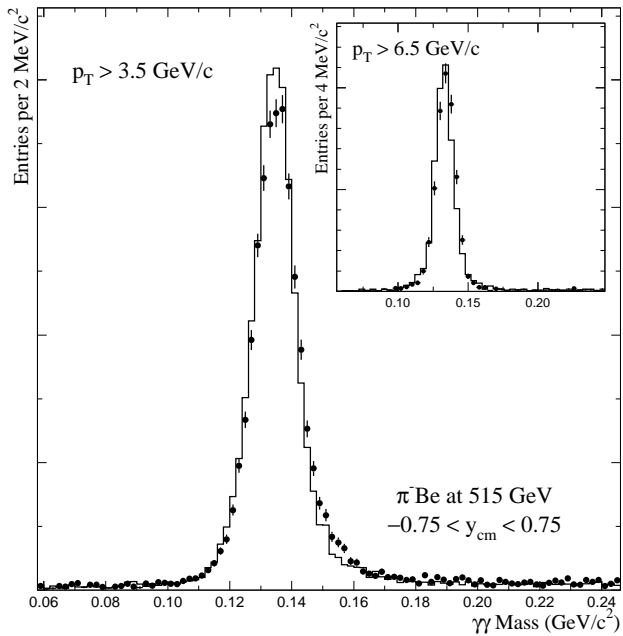}
\caption{Comparison between data~($\bullet$) and the Monte Carlo
(histogram) for $\gamma\gamma$ combinations in the $\pi^0$ mass region 
for two requirements on minimum $p_T$. The distributions have been 
normalized to unit area.}
\label{fig:mc-pi0-eta}
\end{figure}

\begin{figure}
\epsfxsize=\figsize \epsfbox[\bbcoordb]{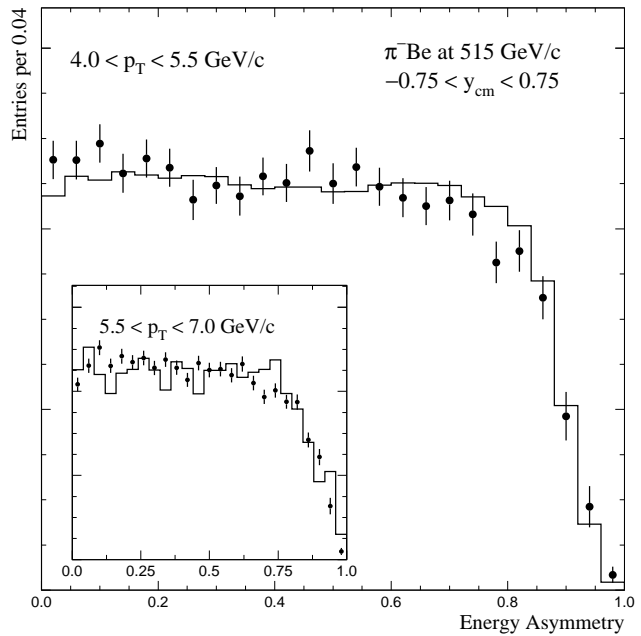}
\caption{Comparison of the background subtracted $\gamma\gamma$ energy 
asymmetry distribution in the $\pi^0$ signal region in data
($\bullet$) and the Monte Carlo (histogram). Shown are the comparisons
for two $p_T$ intervals, $4.0<p_T< 5.5$~GeV/c and $5.5<p_T<7.0$~GeV/c.
The distributions have been normalized to unit area.}
\label{fig:mc-A}
\end{figure}

Figure~\ref{fig:mc-pi0-eta} compares the $\gamma\gamma$~mass spectra 
in the $\pi^0$ and $\eta$ mass regions to the simulation for two different
minimum $p_T$ cutoffs.  In addition to providing evidence that the Monte 
Carlo simulated the EMLAC resolution well, the agreement between the levels 
of combinatorial background also indicates that the Monte Carlo provided a 
reasonable simulation of the underlying event structure.  
Figure~\ref{fig:mc-A} shows a
comparison between the Monte Carlo simulation and the data for the
background subtracted $\gamma\gamma$ energy asymmetry in the $\pi^0$ 
signal region, for two $p_T$ intervals. This figure
illustrates that the simulation accurately describes the losses of
very low-energy photons.

\subsection{Normalization}

Electronic scalers that counted signals from the beam hodoscope,
interaction counters, and beam hole counters were used to evaluate
the number of beam particles incident on the target. Other scalers
logged the state of the trigger and of components of the data
acquisition system. Information from these scalers was used to
determine the number of beam particles that traversed the spectrometer
when it was ready to record data. This number was corrected for
multiple occupancy in the beam ($\approx 3\%$), the absorption
of beam in the target material ($\approx 6\%$ for the Be target and 
$\approx 3\%$ for $p$), and for the $\mu^-$ content of the beam 
which was measured to be $\approx 0.5\%$~\cite{osborne}.

The normalization of the low \pt\ \piz\ cross section was
independently verified using events from the prescaled beam and
interaction trigger samples.  In these samples, the absolute
normalization is obtained simply by counting the recorded
events selected by these triggers. For these low
\pt\ events, the normalization as determined via the scalers and via
event counting techniques agreed to $3\%$ accuracy.

An analysis of negative secondary beam production by 800~GeV/$c$
primary protons indicates a small ($\approx 1\%$) \kmi\ component 
in the incident 515 GeV/$c$ beam. 
The meson production cross sections were corrected for this \kmi\ component
under the assumption that meson production by \kmi\ beam is half that 
of \pim\ beam \cite{striley,donaldson_pi0}.

\begin{figure}
\leavevmode
\epsfxsize=\figsize \epsfbox[\bbcoord]{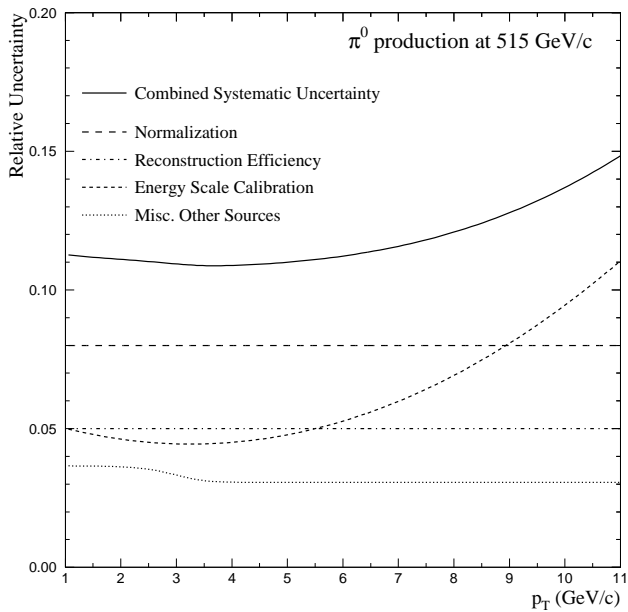}
\caption{Combined relative systematic uncertainty for \piz\ production
as a function of $p_T$ (solid line). Also shown are the contributions
from the various sources of systematic uncertainty.}
\label{fig:pixs_sys_515}
\end{figure}

\subsection{Summary of systematic uncertainties}

The principal contributions to the systematic uncertainty arose from
the following sources: calibration of photon energy response, $\pi^0$
and $\eta$ reconstruction efficiency and detector-resolution
unsmearing, and overall normalization.  The relative systematic
uncertainty for \piz\ production is shown as a function of \pt\ in
Fig.~\ref{fig:pixs_sys_515}. Included in the figure are curves showing the
contributions from the major sources of systematic uncertainty. Other
sources of uncertainty which contribute at the \hbox{$1$--$2\%$} level
include: background subtraction, incident beam contamination, beam halo muon
rejection, geometric acceptance, photon conversions, trigger response,
and vertex finding.  The total systematic uncertainty is calculated by
combining in quadrature all the individual uncertainties.  The
uncertainties for $\eta$ production are similar to those for \piz\
production, except for the uncertainty in the trigger response, which
is $\approx 5\%$ at low \pt.  The actual systematic uncertainties are
quoted in the appropriate tables of cross sections, except at low \pt,
where, the statistical and systematic uncertainties have been combined
because of the large correlation between them.

\begin{figure}
\leavevmode
\epsfxsize=\figsize \epsfbox[\bbcoord]{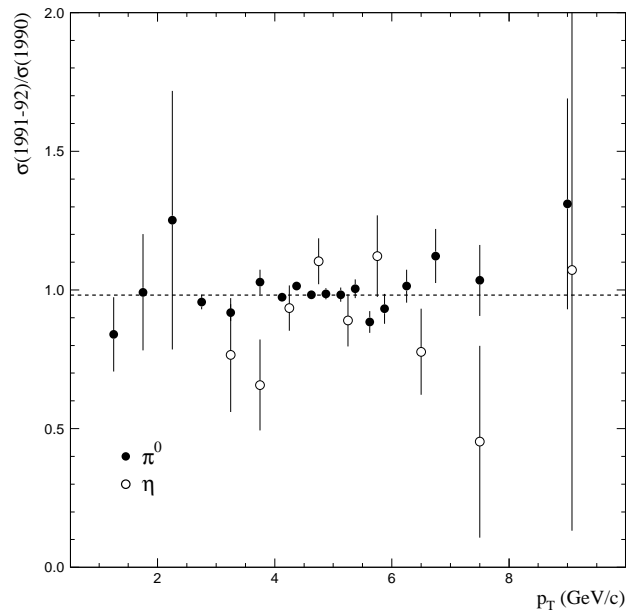}
\caption{Ratio of production cross sections by 515 GeV/$c$
$\pi^-$ beam on Be from the 1991-92 fixed target run to those obtained from 
the 1990 run. The error bars reflect only statistical uncertainties for 
\pt\ $>$ 2.0 GeV/c. The dashed line is a fit to the ratio of $\pi^0$ cross 
sections.}
\label{fig:1991_1990_rat}
\end{figure}

A cross-check of the systematics can be obtained by comparing the
results for meson production from the 1991-92 data sample to the
results from the 1990 data sample for independent analyses. The 
ratio of the $\pi^0$ and $\eta$
production cross sections from the 1991-92 run to the 1990 run is
shown in Fig.~\ref{fig:1991_1990_rat}.  A linear fit to the ratio of
\piz\ cross sections yields $0.982~\pm~0.006$, which is well within
the systematic uncertainties described earlier in this section.

The secondary pion beams were determined to have a mean momentum of
$515~\pm~2~{\rm GeV}/c$ with an estimated halfwidth of $\approx
30~{\rm GeV}/c$. This momentum spread introduces a small uncertainty
($\approx 5\%$) in comparisons of theory with data.

\section{Results and Discussion}

\subsection{\piz\ production}

\begin{figure}[t]
\epsfxsize=\figsize \epsfbox[\bbcoord]{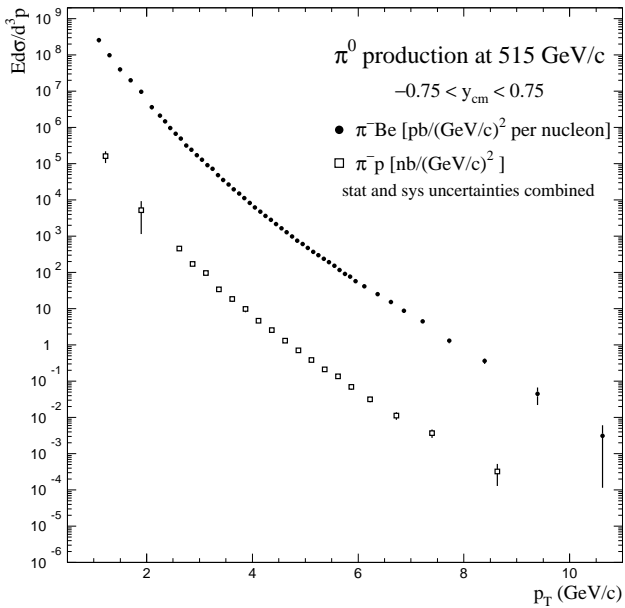}
\caption{
Invariant differential cross sections (per nucleon) for $\pi^0$
production as a function of $p_T$ in $\pi^-p$ and $\pi^-Be$
interactions at 515 GeV/$c$.  Cross sections have been averaged over
the rapidity range, $-0.75 \le \ycm\ \le 0.75$. The error bars
represent the combined statistical and systematic uncertainties.}
\label{fig:pixs_pt_515}
\end{figure}

The inclusive $\pi^0$ cross sections per nucleon versus $p_T$ are
shown in Fig.~\ref{fig:pixs_pt_515} for 515~GeV/$c$ $\pi^-$ beams
incident upon beryllium and liquid hydrogen targets. Note that the
cross section is in units of pb/(GeV/c)$^2$ per nucleon for the Be target, 
and in nb/(GeV/c)$^2$ for the $p$ target. The measurements
on beryllium were obtained using the combined statistics from the 
1990 and 1991-92 runs. Because of the
steeply falling spectra, the data are plotted at abscissa values that
correspond to the average values of the cross section in each $p_T$
bin, assuming local exponential $p_T$ dependence~\cite{laff}. The
cross sections are also tabulated in Tables~\ref{pi0_table_be}
through~\ref{pi0_table_p_rap}.

\begin{figure}[t]
\epsfxsize=\figsize \epsfbox[\bbcoord]{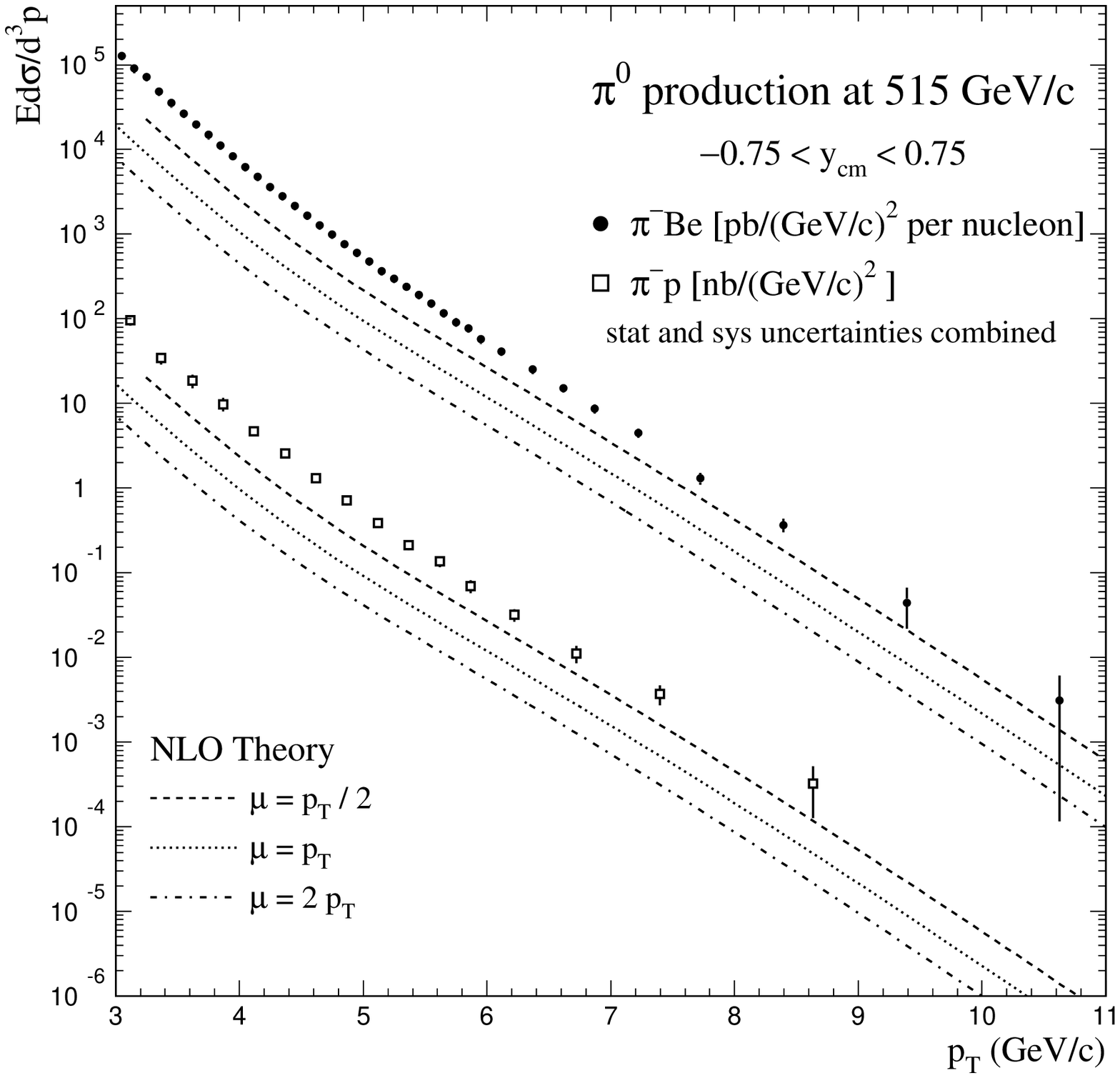}
\caption{
Invariant differential cross sections (per nucleon) for $\pi^0$
production as a function of $p_T$ in $\pi^-p$ and $\pi^-Be$
interactions at 515 GeV/$c$.  Cross sections have been averaged over
the rapidity range, $-0.75 \le \ycm\ \le 0.75$. The error bars
have statistical and systematic uncertainties added in quadrature.
Overlayed on the data are NLO PQCD results.}
\label{fig:pixs_pt_515_qcd}
\end{figure}

In Fig.~\ref{fig:pixs_pt_515_qcd}, the measured inclusive $\pi^0$
cross sections are compared to NLO PQCD results~\cite{aversa} using
GRV~\cite{grv} parton distributions, KKP~\cite{kkp} fragmentation
functions, and factorization scales of $\mu=p_T/2$, $\pt$, and $2\pt$
(the renormalization and fragmentation scales have been set equal to
the value of the factorization scale). The PQCD calculations for the
Be target were adjusted to account for nuclear effects using the {\sc
hijing} Monte Carlo calculation~\cite{wang-note}.
The large scale dependence in the calculations is insufficient to raise the
predicted cross sections up to the values indicated by the data.

These discrepancies have been interpreted~\cite{prl,prd} as arising
from additional soft-gluon emission in the initial state that is not
included in the NLO calculation, and which results in sizeable parton
$k_T$ before the hard collision (for a different perspective, see the
discussion in Ref.~\cite{frenchpiz}).  Soft-gluon (or $k_T$) effects
are expected in all hard-scattering processes, such as the inclusive
production of jets, high-$p_T$ mesons, and direct
photons~\cite{FF,font,cont1,cont2}.  The Collins-Soper-Sterman
resummation formalism~\cite{css} provides a rigorous basis for
understanding these radiative effects, and there have been several
recent efforts to derive resummation descriptions for the inclusive
direct-photon~\cite{Catani:1999hs,Catani,Lai:1998xq,sterman-res,fink},
jet \cite{kidonakis-owens-jet}, and
dijet cross sections~\cite{Laenen,Kidonakis,Kidonakis2}.  The
calculation of Ref.~\cite{Catani:1999hs} for inclusive direct-photon
production, which includes the effects of soft-gluon resummation near
the kinematic threshold limit $\left( x_T = 2 p_T/ \sqrt{s}
\longrightarrow 1 \right)$, has a far smaller sensitivity to scale,
compared to NLO calculations, and provides cross sections close to
those of NLO calculations with a scale of $\mu = p_T/2$.  Also, for
our energies, the calculations of Ref.~\cite{sterman-res,fink}, which 
simultaneously treat threshold and recoil effects in direct-photon
production, yield a substantially larger cross section than the NLO
result.  However, no such calculations are available for inclusive
meson production.  In their absence, we use a PQCD-based model that
incorporates transverse kinematics of initial-state partons to study
the consequences of additional $k_T$ for high-$p_T$ production
processes.

\begin{figure}
\epsfxsize=\figsize \epsfbox[\bbcoord]{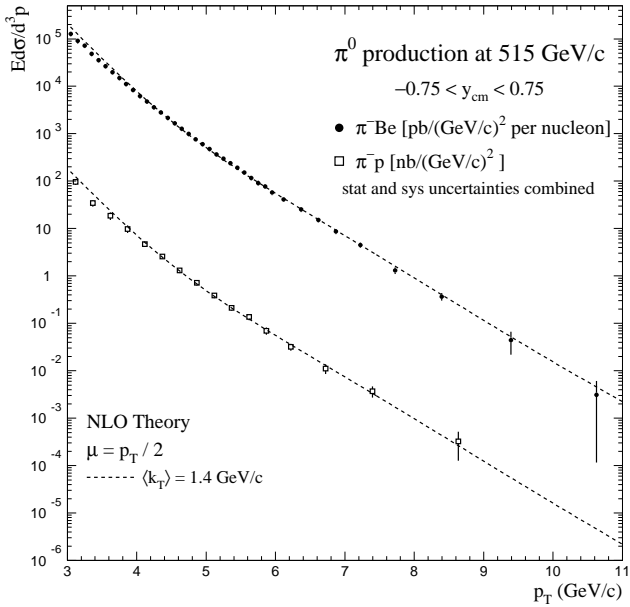}
\caption{
Invariant differential cross sections (per nucleon) for $\pi^0$
production as a function of $p_T$ in $\pi^-p$ and $\pi^-Be$
interactions at 515 GeV/$c$.  Cross sections have been averaged over
the rapidity range, $-0.75 \le \ycm\ \le 0.75$.  The error bars
have statistical and systematic uncertainties added in
quadrature. The curves represent the $k_T$-enhanced NLO QCD
calculations for \avkt=1.4 GeV/$c$.}
\label{fig:pixs_pt_515_kt}
\end{figure}

Because the unmodified PQCD cross sections fall rapidly with
increasing $p_T$, the net effect of the ``$k_T$ smearing'' is to
increase the expected yield. Modified parton kinematics have been 
implemented in a Monte Carlo calculation of the
leading-order (LO) cross sections for high-$p_T$ particle
production~\cite{owensll}, with the $k_T$ distribution for each of the
incoming partons represented by a Gaussian with one adjustable
parameter (\avkt).  Unfortunately, no such program is available for
NLO calculations, and so we approximate the effect of $k_T$ smearing
by multiplying the NLO cross sections by the corresponding LO
$k_T$-enhancement factors.  Admittedly, this procedure involves a risk
of double-counting since some of the $k_T$-enhancement may already be
contained in the NLO calculation. However, we expect such
double-counting effects to be small.

In the calculation of the LO $k_T$-enhancement factors we employ
\avkt\ values representative of those found from comparisons of
kinematic distributions in data involving production of high-mass
$\gamma\gamma$, $\gamma\pi^0$, and $\pi^0\pi^0$ systems with
calculations relying on the same LO program (see Refs.~\cite{prl,prd}
for further details). For these comparisons, we used the LO versions
of the GRV parton distributions and (where appropriate) KKP
fragmentation functions, and an average transverse momentum of
0.6~GeV/$c$ for the $\pi^0$ mesons relative to the fragmenting parton
direction (varying this parameter in the range 0.3--0.7 GeV/$c$ does
not affect our conclusions)~\cite{R702-jets,CCOR-qt}.

\begin{figure}
\epsfxsize=\figsize \epsfbox[\bbcoord]{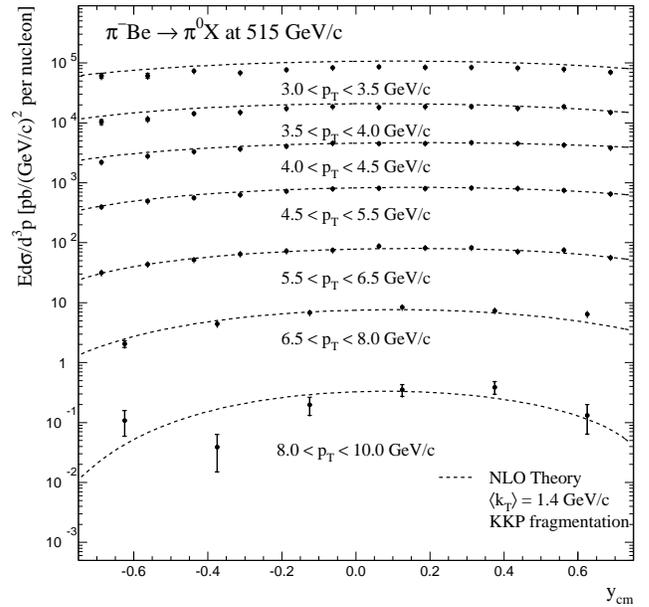}
\caption{
Invariant cross sections per nucleon for $\pi^0$ production in
$\pi^-Be$ interactions at 515~GeV/$c$. Cross sections are shown versus
\ycm\ for several intervals in $p_T$.  The curves represent
$k_T$-enhanced NLO QCD calculations for \avkt=1.4 GeV/$c$. The error bars 
have statistical and systematic uncertainties added in quadrature.}
\label{fig:pixs_rap_515}
\end{figure}

\begin{figure}[!ht]
\epsfxsize=\figsize \epsfbox[\bbcoorde]{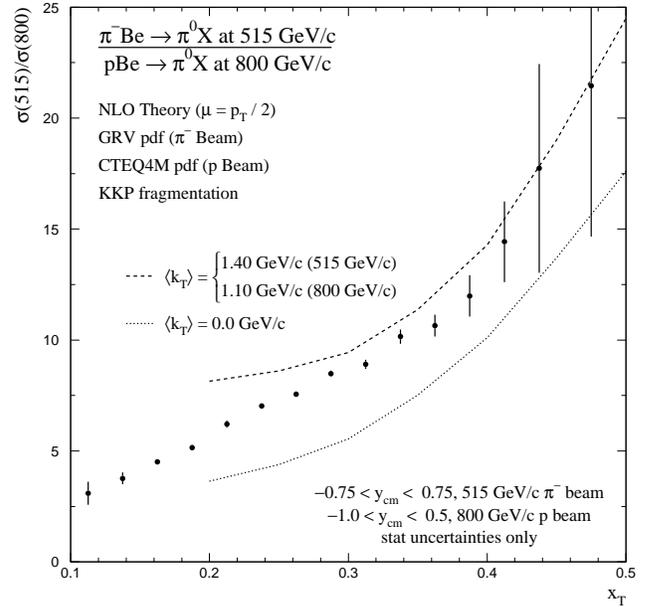}
\caption{
Ratio of 515~GeV/$c$ \pim\ to 800~GeV/$c$ $\mit p$~beam \piz\ production 
cross sections as a function of \xt, compared to conventional 
and \kt-enhanced NLO QCD results.}
\label{fig:xsrat_800_515}
\end{figure}

Comparisons of the $k_T$-enhanced calculations with data at
515~GeV/$c$ are displayed in Fig.~\ref{fig:pixs_pt_515_kt}, indicating
good agreement for the chosen \avkt\ value.
Figure~\ref{fig:pixs_rap_515} shows the $\pi^-Be$ cross sections at
515~GeV/$c$ versus rapidity, for several intervals in $p_T$. The
shapes and normalizations of the data are in good agreement with the
$k_T$-enhanced calculations.

In a previous publication~\cite{E706-pi-eta}, this experiment reported
results for \piz\ production in \pp\ and \pBe\ interactions at 800 and
530 GeV/${c}$. This offers the opportunity to compare production cross
sections by incident \pim\ and $\mit p$ beams.  In
Fig.~\ref{fig:xsrat_800_515}, the ratio of invariant cross sections
for \piz\ production by 515~GeV/$c$ \pim\ and 800~GeV/$c$ $\mit
p$~beam is shown as a function of \xt\ (comparing results at
approximately the same incident momentum per valence quark). Both
theoretical and experimental uncertainties are reduced in the ratio
allowing, in principle, a more sensitive test of the calculations. In
the figure, the data are compared to conventional (\avkt=0) and
\kt-enhanced NLO results using KKP fragmentation functions. The
\kt-enhanced theory accommodates the data better 
than the conventional theory.

\subsection{$\eta$ production}

\begin{figure}[t]
\epsfxsize=\figsize \epsfbox[\bbcoord]{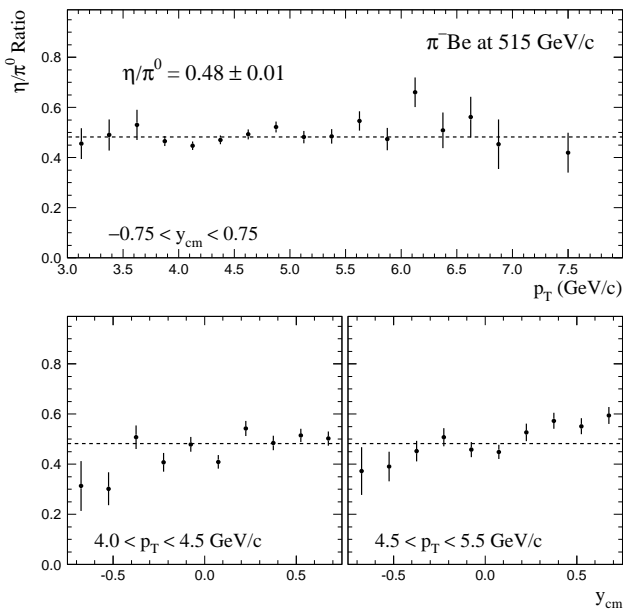}
\caption{Ratios of  $\eta$ to $\pi^0$ invariant cross sections
at 515 GeV/$c$, as functions of $p_T$ (top), and of \ycm\ (bottom)
for two $p_T$ ranges.  The error bars reflect only the statistical
uncertainties.}
\label{fig:etapi_515}
\end{figure}

Cross sections for inclusive $\eta$ production are tabulated in
Tables~\ref{eta_table_be} through \ref{eta_table_p_rap}.  Theoretical
descriptions of $\eta$-meson production differ from the $\pi^0$ case
primarily because of differences in the fragmentation of partons into
the particles of interest.  To investigate this effect, we present
$\eta/\pi^0$ relative production rates as functions of $p_T$ and
\ycm\ (for two $p_T$ ranges) in Fig.~\ref{fig:etapi_515}. We see no
significant dependence in this ratio --- the data average to a value
of $0.48 \pm 0.01$ (statistical). 

\section{Summary}

The invariant cross sections for $\pi^0$ and $\eta$ production have
been measured for $\pi^-p$ and $\pi^-Be$ collisions at 515~GeV/$c$ as
functions of $p_T$ and \ycm, over the kinematic range 
$1<p_T<11$~GeV/$c$, and $-0.75 \le \ycm\ \le 0.75$. The 
$\pi^0$ cross
sections agree with $k_T$-enhanced NLO QCD calculations.  The measured
$\eta/\pi^0$ production ratio, which provides information about the
relative fragmentation of partons into these mesons, is $0.48 \pm 0.01$
with little dependence on $p_T$ or rapidity.

\smallskip
\begin{acknowledgments}

We thank the U.~S. Department of Energy, the National Science
Foundation, including its Office of International Programs, and the
Universities Grants Commission of India, for their support of this
research.  The staff and management of Fermilab are thanked for their
efforts in making available the beam and computing facilities that
made this work possible.  We are also pleased to acknowledge the
contributions of our colleagues on Fermilab experiment E672. We
acknowledge the contributions of the following colleagues for their
help in the operation and upgrade of the Meson West spectrometer:
W.~Dickerson, E.~Pothier from Northeastern University;
J.~T.~Anderson,
E.~Barsotti~Jr.,
H.~Koecher,
P.~Madsen,
D.~Petravick,
R.~Tokarek, J.~Tweed, D.~Allspach, J.~Urbin, and the cryo crews
from Fermi National Accelerator Laboratory;
T.~Haelen, C.~Benson, 
L.~Kuntz, and D.~Ruggiero
from the University of Rochester;
the technical staffs of 
Michigan State University and
Pennsylvania State University for the construction of the straw tubes
and of the University of Pittsburgh for the silicon detectors. We
would also like to thank the following commissioning run collaborators 
for their invaluable contributions to the hardware and software
infrastructure of the original Meson West spectrometer: 
G.~Alverson,
G.~Ballocchi,
R.~Benson,
D.~Berg,
D.~Brown,
D.~Carey,
T.~Chand,
C.~Chandlee,
S.~Easo,
W.~Faissler,
G.~Glass,
I.~Kourbanis,
A.~Lanaro,
C.~Nelson~Jr.,
D.~Orris,
B.~Rajaram,
K.~Ruddick,
A.~Sinanidis, and
G.~Wu.
We thank S.~Catani, J.Ph.~Guillet, B.~Kniehl, J.~Owens, 
G.~Sterman, W.~Vogelsang, and X.~-N.~Wang for many helpful
discussions and for providing us with their QCD calculations.

\end{acknowledgments}


\begin{thebibliography}{53}
\bibitem[{E67()}]{E672-spec}
\bibinfo{note}{The figure does not include the Cherenkov detector, steel
  absorber, and the veto walls located upstream of the detector elements shown.
  The figure also excludes the muon identification system, which was located
  downstream of the detector elements shown in the diagram, and which was used
  primarily by Fermilab E672 to select events containing high mass dimuons.}
\end{thebibliography}

\newpage

\appendix*

\section{Tabulated Cross Sections}

\squeezetable
\begin{table}[h]
\caption{Invariant differential cross section
$\left( \DIFFXS \right)$ per nucleon for the inclusive reaction
\piBe~$\rightarrow\pi^0$X at 515 GeV/$c$, averaged over the rapidity
interval $-0.75 \le \ycm\ \le  0.75$.}
\begin{tabular*}{3.25in}{r@{ -- }l
@{\extracolsep{\fill}}
r@{\extracolsep{1pt}${}\pm{}$}l}
 \hline
 \hline
\multicolumn{2}{c}{$p_T$}  &\multicolumn{2}{c}{${\mit \pi^{-}}$Be @ 515~GeV/$c$} \\
\multicolumn{2}{c}{(GeV/$c$)}  &\multicolumn{2}{c}{$(\rm{\mu{b}}$/(GeV/$c)^2$)} \\
 \hline
  1.00 & 1.20 & 
255 & 35 \\
  1.20 & 1.40 & 
97 & 14 \\
  1.40 & 1.60 & 
39.5 & 5.9 \\
  1.60 & 1.80 & 
19.8 & 2.5 \\
  1.80 & 2.00 & 
9.5 & 1.2 \\
  2.00 & 2.20 & 
3.64 & 0.36${}\pm{}$0.43 \\
  2.20 & 2.30 & 
2.112 & 0.053${}\pm{}$0.25 \\
  2.30 & 2.40 & 
1.460 & 0.038${}\pm{}$0.17 \\
 \hline
\multicolumn{2}{c}{ }  &\multicolumn{2}{c}{$(\rm{nb}$/(GeV/$c)^2$)} \\
 \hline
  2.40 & 2.50 & 
955 & 27${}\pm{}$111 \\
  2.50 & 2.60 & 
673 & 17${}\pm{}$78 \\
  2.60 & 2.70 & 
496 & 14${}\pm{}$57 \\
  2.70 & 2.80 & 
315.4 & 9.6${}\pm{}$36.1 \\
  2.80 & 2.90 & 
243.8 & 7.2${}\pm{}$27.8 \\
  2.90 & 3.00 & 
170.2 & 4.3${}\pm{}$19.3 \\
  3.00 & 3.10 & 
127.5 & 3.3${}\pm{}$14.4 \\
  3.10 & 3.20 & 
90.7 & 2.6${}\pm{}$10.2 \\
  3.20 & 3.30 & 
72.1 & 1.9${}\pm{}$8.1 \\
  3.30 & 3.40 & 
48.3 & 1.3${}\pm{}$5.4 \\
  3.40 & 3.50 & 
35.38 & 1.00${}\pm{}$3.9 \\
  3.50 & 3.60 & 
26.53 & 0.65${}\pm{}$2.9 \\
  3.60 & 3.70 & 
19.73 & 0.50${}\pm{}$2.2 \\
  3.70 & 3.80 & 
14.88 & 0.36${}\pm{}$1.6 \\
  3.80 & 3.90 & 
11.15 & 0.30${}\pm{}$1.2 \\
  3.90 & 4.00 & 
8.28 & 0.11${}\pm{}$0.91 \\
  4.00 & 4.10 & 
6.220 & 0.039${}\pm{}$0.69 \\
  4.10 & 4.20 & 
4.770 & 0.034${}\pm{}$0.53 \\
  4.20 & 4.30 & 
3.604 & 0.027${}\pm{}$0.40 \\
  4.30 & 4.40 & 
2.825 & 0.022${}\pm{}$0.31 \\
  4.40 & 4.50 & 
2.166 & 0.019${}\pm{}$0.24 \\
  4.50 & 4.60 & 
1.654 & 0.015${}\pm{}$0.18 \\
  4.60 & 4.70 & 
1.277 & 0.013${}\pm{}$0.14 \\
 \hline
\multicolumn{2}{c}{ }  &\multicolumn{2}{c}{$(\rm{pb}$/(GeV/$c)^2$)} \\
 \hline
  4.70 & 4.80 & 
990 & 11${}\pm{}$108 \\
  4.80 & 4.90 & 
756.4 & 9.5${}\pm{}$83.0 \\
  4.90 & 5.00 & 
605.8 & 8.5${}\pm{}$66.6 \\
  5.00 & 5.10 & 
475.9 & 7.2${}\pm{}$52.4 \\
  5.10 & 5.20 & 
366.0 & 6.0${}\pm{}$40.3 \\
  5.20 & 5.30 & 
298.1 & 5.4${}\pm{}$32.9 \\
  5.30 & 5.40 & 
237.8 & 4.8${}\pm{}$26.3 \\
  5.40 & 5.50 & 
191.4 & 4.3${}\pm{}$21.2 \\
  5.50 & 5.60 & 
151.3 & 3.7${}\pm{}$16.8 \\
  5.60 & 5.70 & 
116.5 & 3.2${}\pm{}$13.0 \\
  5.70 & 5.80 & 
91.0 & 2.8${}\pm{}$10.2 \\
  5.80 & 5.90 & 
77.2 & 2.6${}\pm{}$8.6 \\
  5.90 & 6.00 & 
57.3 & 2.2${}\pm{}$6.4 \\
  6.00 & 6.25 & 
41.3 & 1.2${}\pm{}$4.7 \\
  6.25 & 6.50 & 
25.27 & 0.90${}\pm{}$2.9 \\
  6.50 & 6.75 & 
15.27 & 0.67${}\pm{}$1.7 \\
  6.75 & 7.00 & 
8.71 & 0.50${}\pm{}$1.0 \\
  7.00 & 7.50 & 
4.48 & 0.25${}\pm{}$0.52 \\
  7.50 & 8.00 & 
1.30 & 0.13${}\pm{}$0.16 \\
  8.00 & 9.00 & 
0.366 & 0.049${}\pm{}$0.045 \\
   9.00 & 10.00 & 
0.044 & 0.022${}\pm{}$0.006 \\
  10.00 & 12.00 & 
0.0031 & 0.0031${}\pm{}$0.0004 \\
 \hline
 \hline
 \end{tabular*}
\label{pi0_table_be}
\end{table}

\squeezetable
\begin{table}
\caption{Invariant differential cross section
$\left( \DIFFXS \right)$ for the inclusive reaction
\pip~$\rightarrow\pi^0$X at 515 GeV/$c$, averaged over the rapidity
interval $-0.75 \le \ycm\ \le 0.75$.}
\squeezetable
\begin{tabular*}{3.25in}{r@{ -- }l
@{\extracolsep{\fill}}
r@{\extracolsep{1pt}${}\pm{}$}l}
 \hline
 \hline
\multicolumn{2}{c}{$p_T$}  &\multicolumn{2}{c}{${\mit \pi^{-}p}$ @  515~GeV/$c$} \\
\multicolumn{2}{c}{(GeV/$c$)}  &\multicolumn{2}{c}{$(\rm{\mu{b}}$/(GeV/$c)^2$)} \\
 \hline
  1.00 & 1.50 & 
161 & 57 \\
  1.50 & 2.50 & 
5.2 & 4.1 \\
 \hline
\multicolumn{2}{c}{ }  &\multicolumn{2}{c}{$(\rm{nb}$/(GeV/$c)^2$)} \\
 \hline
  2.50 & 2.75 & 
455 & 34${}\pm{}$52 \\
  2.75 & 3.00 & 
170 & 14${}\pm{}$19 \\
  3.00 & 3.25 & 
96.3 & 8.3${}\pm{}$10.9 \\
  3.25 & 3.50 & 
34.3 & 3.7${}\pm{}$3.8 \\
  3.50 & 3.75 & 
18.5 & 2.6${}\pm{}$2.1 \\
  3.75 & 4.00 & 
9.8 & 1.4${}\pm{}$1.1 \\
  4.00 & 4.25 & 
4.67 & 0.12${}\pm{}$0.51 \\
  4.25 & 4.50 & 
2.586 & 0.085${}\pm{}$0.28 \\
  4.50 & 4.75 & 
1.303 & 0.048${}\pm{}$0.14 \\
 \hline
\multicolumn{2}{c}{ }  &\multicolumn{2}{c}{$(\rm{pb}$/(GeV/$c)^2$)} \\
 \hline
  4.75 & 5.00 & 
715 & 33${}\pm{}$79 \\
  5.00 & 5.25 & 
386 & 24${}\pm{}$43 \\
  5.25 & 5.50 & 
213 & 16${}\pm{}$24 \\
  5.50 & 5.75 & 
137 & 13${}\pm{}$15 \\
  5.75 & 6.00 & 
69.7 & 9.3${}\pm{}$7.8 \\
  6.00 & 6.50 & 
31.8 & 4.1${}\pm{}$3.6 \\
  6.50 & 7.00 & 
11.2 & 2.3${}\pm{}$1.3 \\
  7.00 & 8.00 & 
3.69 & 0.86${}\pm{}$0.43 \\
   8.00 & 10.00 & 
0.32 & 0.19${}\pm{}$0.04 \\
 \hline
 \hline
 \end{tabular*}
\label{pi0_table_p}
\end{table}

\newpage
\clearpage
\begin{widetext}
\widetext
\begin{table}
\caption{The averaged invariant differential cross section 
$\left( \DIFFXS \right)$ per nucleon as a function of rapidity and $p_T$
for the inclusive reaction \piBe~$\rightarrow\pi^0$X at
515~GeV/$c$.}
\begin{tabular}{r@{ -- }l
r@{ }l
r@{ }l
r@{ }l
r@{ }l}
 \hline
 \hline
\multicolumn{2}{c}{$y_{cm}$}  &\multicolumn{8}{c}{$p_T$ (GeV/$c$)} \\
\multicolumn{2}{c}{ } &\multicolumn{2}{c}{1.00{ -- }1.50 } &\multicolumn{2}{c}{1.50{ -- }2.00 } &\multicolumn{2}{c}{2.00{ -- }2.50 } &\multicolumn{2}{c}{2.50{ -- }3.00 } \\
\multicolumn{2}{c}{ } &\multicolumn{2}{c}{$(\rm{\mu{b}}$/(GeV/$c)^2)$} &\multicolumn{2}{c}{$(\rm{\mu{b}}$/(GeV/$c)^2)$} &\multicolumn{2}{c}{$(\rm{\mu{b}}$/(GeV/$c)^2)$} &\multicolumn{2}{c}{$(\rm{nb}$/(GeV/$c)^2)$} \\
 \hline         
  $-0.750$ & $-0.625$ & 
\vctr{132${\,}\pm$} & \vctr{30} & \vctr{13.3${\,}\pm$} & \vctr{3.7} & \vctr{1.84${\,}\pm$} & \vctr{0.86${}\pm{}$0.22} & {373${\,}\pm$} & {26${}\pm{}$43} \\
  $-0.625$ & $-0.500$ & 
   &    &    &    &   &  & {400${\,}\pm$} & {24${}\pm{}$46} \\
  $-0.500$ & $-0.375$ & 
\vctr{175${\,}\pm$} & \vctr{29} & \vctr{18.4${\,}\pm$} & \vctr{4.4} & {2.63${\,}\pm$} & {0.13${}\pm{}$0.31} & {366${\,}\pm$} & {24${}\pm{}$42} \\
  $-0.375$ & $-0.250$ & 
   &    &    &    & {2.48${\,}\pm$} & {0.10${}\pm{}$0.29} & {336${\,}\pm$} & {16${}\pm{}$38} \\
  $-0.250$ & $-0.125$ & 
\vctr{148${\,}\pm$} & \vctr{26} & \vctr{23.2${\,}\pm$} & \vctr{4.0} & {2.538${\,}\pm$} & {0.084${}\pm{}$0.30} & {375${\,}\pm$} & {16${}\pm{}$43} \\
  $-0.125$ & $0.000$  & 
   &    &    &    & {2.522${\,}\pm$} & {0.068${}\pm{}$0.30} & {404${\,}\pm$} & {15${}\pm{}$46} \\
  $0.000$  & $0.125$  & 
\vctr{171${\,}\pm$} & \vctr{31} & \vctr{19.6${\,}\pm$} & \vctr{4.0} & {2.506${\,}\pm$} & {0.064${}\pm{}$0.29} & {409${\,}\pm$} & {13${}\pm{}$47} \\
  $0.125$  & $0.250$  & 
   &    &    &    & {2.551${\,}\pm$} & {0.063${}\pm{}$0.30} & {392${\,}\pm$} & {12${}\pm{}$45} \\
  $0.250$  & $0.375$  & 
\vctr{124${\,}\pm$} & \vctr{22} & \vctr{24.6${\,}\pm$} & \vctr{4.4} & {2.451${\,}\pm$} & {0.067${}\pm{}$0.29} & {393${\,}\pm$} & {12${}\pm{}$45} \\
  $0.375$  & $0.500$  & 
   &    &    &    & {2.360${\,}\pm$} & {0.070${}\pm{}$0.28} & {354${\,}\pm$} & {11${}\pm{}$41} \\
  $0.500$  & $0.625$  & 
\vctr{144${\,}\pm$} & \vctr{24} & \vctr{14.8${\,}\pm$} & \vctr{3.2} & {2.407${\,}\pm$} & {0.074${}\pm{}$0.28} & {368${\,}\pm$} & {12${}\pm{}$42} \\
  $0.625$  & $0.750$  & 
   &    &    &    & {2.083${\,}\pm$} & {0.071${}\pm{}$0.24} & {334${\,}\pm$} & {12${}\pm{}$38} \\
 \hline         
\multicolumn{2}{c}{ } &\multicolumn{2}{c}{3.00{ -- }3.50 } &\multicolumn{2}{c}{3.50{ -- }4.00 } &\multicolumn{2}{c}{4.00{ -- }4.50 } &\multicolumn{2}{c}{4.50{ -- }5.00 } \\
\multicolumn{2}{c}{ } &\multicolumn{2}{c}{$(\rm{nb}$/(GeV/$c)^2)$} &\multicolumn{2}{c}{$(\rm{nb}$/(GeV/$c)^2)$} &\multicolumn{2}{c}{$(\rm{nb}$/(GeV/$c)^2)$} &\multicolumn{2}{c}{$(\rm{nb}$/(GeV/$c)^2)$} \\
 \hline         
  $-0.750$ & $-0.625$ & 
{60.8${\,}\pm$} & {5.7${}\pm{}$6.8} & {10.38${\,}\pm$} & {0.97${}\pm{}$1.1} & {2.198${\,}\pm$} & {0.071${}\pm{}$0.24} & {0.614${\,}\pm$} & {0.022${}\pm{}$0.067} \\
  $-0.625$ & $-0.500$ & 
{61.0${\,}\pm$} & {5.8${}\pm{}$6.8} & {11.59${\,}\pm$} & {0.95${}\pm{}$1.3} & {2.766${\,}\pm$} & {0.057${}\pm{}$0.30} & {0.770${\,}\pm$} & {0.019${}\pm{}$0.084} \\
  $-0.500$ & $-0.375$ & 
{73.3${\,}\pm$} & {4.1${}\pm{}$8.2} & {14.27${\,}\pm$} & {0.46${}\pm{}$1.6} & {3.309${\,}\pm$} & {0.044${}\pm{}$0.36} & {0.863${\,}\pm$} & {0.016${}\pm{}$0.095} \\
  $-0.375$ & $-0.250$ & 
{68.6${\,}\pm$} & {3.3${}\pm{}$7.7} & {14.96${\,}\pm$} & {0.57${}\pm{}$1.7} & {3.668${\,}\pm$} & {0.041${}\pm{}$0.40} & {0.955${\,}\pm$} & {0.017${}\pm{}$0.10} \\
  $-0.250$ & $-0.125$ & 
{77.1${\,}\pm$} & {2.9${}\pm{}$8.7} & {17.31${\,}\pm$} & {0.65${}\pm{}$1.9} & {4.067${\,}\pm$} & {0.038${}\pm{}$0.45} & {1.117${\,}\pm$} & {0.018${}\pm{}$0.12} \\
  $-0.125$ & $0.000$  & 
{83.3${\,}\pm$} & {2.4${}\pm{}$9.3} & {18.56${\,}\pm$} & {0.53${}\pm{}$2.1} & {4.562${\,}\pm$} & {0.040${}\pm{}$0.50} & {1.229${\,}\pm$} & {0.018${}\pm{}$0.13} \\
  $0.000$  & $0.125$  & 
{86.6${\,}\pm$} & {1.7${}\pm{}$9.7} & {18.14${\,}\pm$} & {0.53${}\pm{}$2.0} & {4.535${\,}\pm$} & {0.037${}\pm{}$0.50} & {1.245${\,}\pm$} & {0.017${}\pm{}$0.14} \\
  $0.125$  & $0.250$  & 
{84.6${\,}\pm$} & {1.7${}\pm{}$9.5} & {18.58${\,}\pm$} & {0.60${}\pm{}$2.1} & {4.526${\,}\pm$} & {0.037${}\pm{}$0.50} & {1.218${\,}\pm$} & {0.017${}\pm{}$0.13} \\
  $0.250$  & $0.375$  & 
{84.5${\,}\pm$} & {1.7${}\pm{}$9.5} & {18.63${\,}\pm$} & {0.60${}\pm{}$2.1} & {4.673${\,}\pm$} & {0.040${}\pm{}$0.51} & {1.262${\,}\pm$} & {0.018${}\pm{}$0.14} \\
  $0.375$  & $0.500$  & 
{81.8${\,}\pm$} & {1.8${}\pm{}$9.2} & {17.53${\,}\pm$} & {0.63${}\pm{}$1.9} & {4.531${\,}\pm$} & {0.041${}\pm{}$0.50} & {1.242${\,}\pm$} & {0.019${}\pm{}$0.14} \\
  $0.500$  & $0.625$  & 
{78.0${\,}\pm$} & {1.8${}\pm{}$8.8} & {18.52${\,}\pm$} & {0.58${}\pm{}$2.1} & {4.262${\,}\pm$} & {0.040${}\pm{}$0.47} & {1.159${\,}\pm$} & {0.019${}\pm{}$0.13} \\
  $0.625$  & $0.750$  & 
{70.1${\,}\pm$} & {1.7${}\pm{}$7.9} & {14.96${\,}\pm$} & {0.51${}\pm{}$1.7} & {3.820${\,}\pm$} & {0.039${}\pm{}$0.42} & {1.005${\,}\pm$} & {0.017${}\pm{}$0.11} \\
 \hline         
\multicolumn{2}{c}{ } &\multicolumn{2}{c}{5.00{ -- }5.50 } &\multicolumn{2}{c}{5.50{ -- }6.50 } &\multicolumn{2}{c}{6.50{ -- }8.00 } &\multicolumn{2}{c}{8.00{ -- }10.00} \\
\multicolumn{2}{c}{ } &\multicolumn{2}{c}{$(\rm{pb}$/(GeV/$c)^2)$} &\multicolumn{2}{c}{$(\rm{pb}$/(GeV/$c)^2)$} &\multicolumn{2}{c}{$(\rm{pb}$/(GeV/$c)^2)$} &\multicolumn{2}{c}{$(\rm{pb}$/(GeV/$c)^2)$} \\
 \hline         
  $-0.750$ & $-0.625$ & 
{175.4${\,}\pm$} & {8.4${}\pm{}$19} & {31.5${\,}\pm$} & {2.1${}\pm{}$3.5} & \vctr{2.06${\,}\pm$} & \vctr{0.25${}\pm{}$0.24} & \vctr{0.108${\,}\pm$} & \vctr{0.049${}\pm{}$0.013} \\
  $-0.625$ & $-0.500$ & 
{210.7${\,}\pm$} & {7.4${}\pm{}$23} & {43.7${\,}\pm$} & {2.1${}\pm{}$4.9} &    &    &    &    \\
  $-0.500$ & $-0.375$ & 
{250.9${\,}\pm$} & {8.1${}\pm{}$28} & {51.5${\,}\pm$} & {2.3${}\pm{}$5.8} & \vctr{4.42${\,}\pm$} & \vctr{0.34${}\pm{}$0.51} & \vctr{0.039${\,}\pm$} & \vctr{0.024${}\pm{}$0.005} \\
  $-0.375$ & $-0.250$ & 
{303.5${\,}\pm$} & {8.9${}\pm{}$34} & {64.6${\,}\pm$} & {2.6${}\pm{}$7.2} &    &    &    &    \\
  $-0.250$ & $-0.125$ & 
{323.5${\,}\pm$} & {8.7${}\pm{}$36} & {72.7${\,}\pm$} & {2.8${}\pm{}$8.1} & \vctr{6.76${\,}\pm$} & \vctr{0.43${}\pm{}$0.79} & \vctr{0.197${\,}\pm$} & \vctr{0.066${}\pm{}$0.024} \\
  $-0.125$ & $0.000$  & 
{352.5${\,}\pm$} & {8.9${}\pm{}$39} & {75.0${\,}\pm$} & {2.7${}\pm{}$8.4} &    &    &    &    \\
  $0.000$  & $0.125$  & 
{377.2${\,}\pm$} & {8.9${}\pm{}$42} & {87.6${\,}\pm$} & {2.8${}\pm{}$9.8} & \vctr{8.35${\,}\pm$} & \vctr{0.45${}\pm{}$0.97} & \vctr{0.351${\,}\pm$} & \vctr{0.079${}\pm{}$0.043} \\
  $0.125$  & $0.250$  & 
{379.9${\,}\pm$} & {8.8${}\pm{}$42} & {81.7${\,}\pm$} & {2.7${}\pm{}$9.2} &    &    &    &    \\
  $0.250$  & $0.375$  & 
{383.9${\,}\pm$} & {9.1${}\pm{}$42} & {82.4${\,}\pm$} & {2.8${}\pm{}$9.2} & \vctr{7.33${\,}\pm$} & \vctr{0.45${}\pm{}$0.85} & \vctr{0.389${\,}\pm$} & \vctr{0.092${}\pm{}$0.047} \\
  $0.375$  & $0.500$  & 
{371.3${\,}\pm$} & {9.5${}\pm{}$41} & {71.2${\,}\pm$} & {2.8${}\pm{}$8.0} &    &    &    &    \\
  $0.500$  & $0.625$  & 
{340.1${\,}\pm$} & {9.3${}\pm{}$38} & {75.3${\,}\pm$} & {2.9${}\pm{}$8.4} & \vctr{6.42${\,}\pm$} & \vctr{0.48${}\pm{}$0.75} & \vctr{0.132${\,}\pm$} & \vctr{0.068${}\pm{}$0.016} \\
  $0.625$  & $0.750$  & 
{296.4${\,}\pm$} & {8.8${}\pm{}$33} & {56.3${\,}\pm$} & {2.5${}\pm{}$6.3} &    &    &    &    \\
 \hline
 \hline
 \end{tabular}
\label{pi0_table_be_rap}
\end{table}

\newpage
\clearpage

\begin{table}
\caption{The averaged invariant differential cross section
$\left( \DIFFXS \right)$ as a function of rapidity and $p_T$
for the inclusive reaction \pip~$\rightarrow\pi^0$X at 515~GeV/$c$.}
\squeezetable
\begin{tabular}{r@{ -- }l
r@{ }l
r@{ }l
r@{ }l
r@{ }l}
 \hline
 \hline
\multicolumn{2}{c}{$y_{cm}$}  &\multicolumn{8}{c}{$p_T$ (GeV/$c$)} \\
\multicolumn{2}{c}{ } &\multicolumn{2}{c}{2.50{ -- }3.00 } &\multicolumn{2}{c}{3.00{ -- }3.50 } &\multicolumn{2}{c}{3.50{ -- }4.00 } &\multicolumn{2}{c}{4.00{ -- }4.50 } \\
\multicolumn{2}{c}{ } &\multicolumn{2}{c}{$(\rm{nb}$/(GeV/$c)^2)$} &\multicolumn{2}{c}{$(\rm{nb}$/(GeV/$c)^2)$} &\multicolumn{2}{c}{$(\rm{nb}$/(GeV/$c)^2)$} &\multicolumn{2}{c}{$(\rm{nb}$/(GeV/$c)^2)$} \\
 \hline         
  $-0.750$ & $-0.625$ & 
\vctr{288${\,}\pm$} & \vctr{70${}\pm{}$33} & \vctr{51${\,}\pm$} & \vctr{15${}\pm{}$5.7} & \vctr{5.5${\,}\pm$} & \vctr{3.7${}\pm{}$0.6} & {1.95${\,}\pm$} & {0.29${}\pm{}$0.21} \\
  $-0.625$ & $-0.500$ & 
   &    &    &    &    &    & {2.92${\,}\pm$} & {0.34${}\pm{}$0.32} \\
  $-0.500$ & $-0.375$ & 
\vctr{297${\,}\pm$} & \vctr{62${}\pm{}$34} & \vctr{50.1${\,}\pm$} & \vctr{9.3${}\pm{}$5.6} & \vctr{3.8${\,}\pm$} & \vctr{2.9${}\pm{}$0.4} & {3.55${\,}\pm$} & {0.34${}\pm{}$0.39} \\
  $-0.375$ & $-0.250$ & 
   &    &    &    &    &    & {3.79${\,}\pm$} & {0.28${}\pm{}$0.42} \\
  $-0.250$ & $-0.125$ & 
\vctr{291${\,}\pm$} & \vctr{31${}\pm{}$33} & \vctr{76${\,}\pm$} & \vctr{12${}\pm{}$8.6} & \vctr{16.0${\,}\pm$} & \vctr{3.0${}\pm{}$1.8} & {3.56${\,}\pm$} & {0.29${}\pm{}$0.39} \\
  $-0.125$ & $0.000$  & 
   &    &    &    &    &    & {3.85${\,}\pm$} & {0.21${}\pm{}$0.42} \\
  $0.000$  & $0.125$  & 
\vctr{346${\,}\pm$} & \vctr{30${}\pm{}$40} & \vctr{76${\,}\pm$} & \vctr{10${}\pm{}$8.5} & \vctr{22.5${\,}\pm$} & \vctr{3.9${}\pm{}$2.5} & {4.59${\,}\pm$} & {0.22${}\pm{}$0.51} \\
  $0.125$  & $0.250$  & 
   &    &    &    &    &    & {4.03${\,}\pm$} & {0.20${}\pm{}$0.44} \\
  $0.250$  & $0.375$  & 
\vctr{367${\,}\pm$} & \vctr{25${}\pm{}$42} & \vctr{67.2${\,}\pm$} & \vctr{9.2${}\pm{}$7.5} & \vctr{20.6${\,}\pm$} & \vctr{3.3${}\pm{}$2.3} & {4.19${\,}\pm$} & {0.20${}\pm{}$0.46} \\
  $0.375$  & $0.500$  & 
   &    &    &    &    &    & {3.96${\,}\pm$} & {0.22${}\pm{}$0.44} \\
  $0.500$  & $0.625$  & 
\vctr{286${\,}\pm$} & \vctr{27${}\pm{}$33} & \vctr{72${\,}\pm$} & \vctr{11${}\pm{}$8.1} & \vctr{16.6${\,}\pm$} & \vctr{4.5${}\pm{}$1.8} & {3.86${\,}\pm$} & {0.25${}\pm{}$0.42} \\
  $0.625$  & $0.750$  & 
   &    &    &    &    &    & {3.29${\,}\pm$} & {0.23${}\pm{}$0.36} \\
 \hline         
\multicolumn{2}{c}{ } &\multicolumn{2}{c}{4.50{ -- }5.00 } &\multicolumn{2}{c}{5.00{ -- }5.50 } &\multicolumn{2}{c}{5.50{ -- }6.50 } &\multicolumn{2}{c}{6.50{ -- }8.00 } \\
\multicolumn{2}{c}{ } &\multicolumn{2}{c}{$(\rm{nb}$/(GeV/$c)^2)$} &\multicolumn{2}{c}{$(\rm{pb}$/(GeV/$c)^2)$} &\multicolumn{2}{c}{$(\rm{pb}$/(GeV/$c)^2)$} &\multicolumn{2}{c}{$(\rm{pb}$/(GeV/$c)^2)$} \\
 \hline         
  $-0.750$ & $-0.625$ & 
{0.488${\,}\pm$} & {0.090${}\pm{}$0.054} & {118${\,}\pm$} & {29${}\pm{}$13} & {45${\,}\pm$} & {14${}\pm{}$5.0} & \vctr{3.2${\,}\pm$} & \vctr{1.9${}\pm{}$0.4} \\
  $-0.625$ & $-0.500$ & 
{0.67${\,}\pm$} & {0.11${}\pm{}$0.07} & {259${\,}\pm$} & {67${}\pm{}$29} & {68${\,}\pm$} & {18${}\pm{}$7.6} &    &    \\
  $-0.500$ & $-0.375$ & 
{0.89${\,}\pm$} & {0.10${}\pm{}$0.10} & {316${\,}\pm$} & {54${}\pm{}$35} & {75${\,}\pm$} & {16${}\pm{}$8.4} & \vctr{2.2${\,}\pm$} & \vctr{1.3${}\pm{}$0.3} \\
  $-0.375$ & $-0.250$ & 
{0.892${\,}\pm$} & {0.094${}\pm{}$0.098} & {275${\,}\pm$} & {45${}\pm{}$30} & {70${\,}\pm$} & {17${}\pm{}$7.8} &    &    \\
  $-0.250$ & $-0.125$ & 
{1.116${\,}\pm$} & {0.095${}\pm{}$0.12} & {362${\,}\pm$} & {49${}\pm{}$40} & {61${\,}\pm$} & {14${}\pm{}$6.8} & \vctr{7.4${\,}\pm$} & \vctr{2.5${}\pm{}$0.9} \\
  $-0.125$ & $0.000$  & 
{1.15${\,}\pm$} & {0.10${}\pm{}$0.13} & {404${\,}\pm$} & {54${}\pm{}$45} & {61${\,}\pm$} & {13${}\pm{}$6.9} &    &    \\
  $0.000$  & $0.125$  & 
{1.105${\,}\pm$} & {0.094${}\pm{}$0.12} & {322${\,}\pm$} & {45${}\pm{}$36} & {88${\,}\pm$} & {16${}\pm{}$9.8} & \vctr{9.2${\,}\pm$} & \vctr{2.7${}\pm{}$1.1} \\
  $0.125$  & $0.250$  & 
{1.43${\,}\pm$} & {0.11${}\pm{}$0.16} & {338${\,}\pm$} & {48${}\pm{}$37} & {68${\,}\pm$} & {13${}\pm{}$7.7} &    &    \\
  $0.250$  & $0.375$  & 
{1.274${\,}\pm$} & {0.099${}\pm{}$0.14} & {377${\,}\pm$} & {50${}\pm{}$42} & {102${\,}\pm$} & {17${}\pm{}$11} & \vctr{7.5${\,}\pm$} & \vctr{2.4${}\pm{}$0.9} \\
  $0.375$  & $0.500$  & 
{1.093${\,}\pm$} & {0.098${}\pm{}$0.12} & {357${\,}\pm$} & {51${}\pm{}$39} & {58${\,}\pm$} & {15${}\pm{}$6.5} &    &    \\
  $0.500$  & $0.625$  & 
{1.08${\,}\pm$} & {0.11${}\pm{}$0.12} & {250${\,}\pm$} & {47${}\pm{}$28} & {56${\,}\pm$} & {14${}\pm{}$6.3} & \vctr{7.6${\,}\pm$} & \vctr{2.7${}\pm{}$0.9} \\
  $0.625$  & $0.750$  & 
{0.92${\,}\pm$} & {0.10${}\pm{}$0.10} & {220${\,}\pm$} & {46${}\pm{}$24} & {57${\,}\pm$} & {16${}\pm{}$6.4} &    &    \\
 \hline
 \hline
 \end{tabular}
\label{pi0_table_p_rap}
\end{table}
\end{widetext}

\begin{table}
\caption{Invariant differential cross section
$\left( \DIFFXS \right)$ per nucleon for the inclusive reaction
\piBe~$\rightarrow\eta$X at 515~GeV/$c$, averaged over the rapidity
interval $-0.75 \le \ycm\ \le 0.75$.}
\begin{tabular*}{3.25in}{r@{ -- }l
@{\extracolsep{\fill}}
r@{\extracolsep{1pt}${}\pm{}$}l}
 \hline
 \hline
\multicolumn{2}{c}{$p_T$}  &\multicolumn{2}{c}{${\mit \pi^{-}}$Be @ 515~GeV/$c$} \\
\multicolumn{2}{c}{(GeV/$c$)}  &\multicolumn{2}{c}{$(\rm{nb}$/(GeV/$c)^2$)} \\
 \hline
  3.00 & 3.20 & 
54.5 & 7.5${}\pm{}$6.9 \\
  3.20 & 3.40 & 
23.8 & 3.9${}\pm{}$3.0 \\
  3.40 & 3.60 & 
16.9 & 2.1${}\pm{}$2.1 \\
  3.60 & 3.80 & 
7.91 & 0.97${}\pm{}$0.94 \\
  3.80 & 4.00 & 
4.42 & 0.20${}\pm{}$0.52 \\
  4.00 & 4.20 & 
2.440 & 0.099${}\pm{}$0.28 \\
  4.20 & 4.40 & 
1.506 & 0.062${}\pm{}$0.17 \\
 \hline
\multicolumn{2}{c}{ }  &\multicolumn{2}{c}{$(\rm{pb}$/(GeV/$c)^2$)} \\
 \hline
  4.40 & 4.60 & 
922 & 41${}\pm{}$107 \\
  4.60 & 4.80 & 
571 & 25${}\pm{}$66 \\
  4.80 & 5.00 & 
353 & 16${}\pm{}$41 \\
  5.00 & 5.25 & 
192.9 & 9.5${}\pm{}$22.4 \\
  5.25 & 5.50 & 
110.4 & 6.3${}\pm{}$12.9 \\
  5.50 & 5.75 & 
69.3 & 4.7${}\pm{}$8.1 \\
  5.75 & 6.00 & 
33.4 & 3.0${}\pm{}$3.9 \\
  6.00 & 6.50 & 
20.1 & 1.5${}\pm{}$2.4 \\
  6.50 & 7.00 & 
6.23 & 0.73${}\pm{}$0.75 \\
  7.00 & 8.00 & 
1.21 & 0.22${}\pm{}$0.15 \\
  8.00 & 9.00 & 
0.224 & 0.083${}\pm{}$0.029 \\
   9.00 & 10.00 & 
0.035 & 0.039${}\pm{}$0.005 \\
 \hline
 \hline
 \end{tabular*}
\label{eta_table_be}
\end{table}

\begin{table}
\caption{Invariant differential cross section
$\left( \DIFFXS \right)$ for the inclusive reaction
\pip~$\rightarrow\eta$X at 515~GeV/$c$, averaged over the rapidity
interval $-0.75 \le \ycm\ \le 0.75$.}
\begin{tabular*}{3.25in}{r@{ -- }l
@{\extracolsep{\fill}}
r@{\extracolsep{1pt}${}\pm{}$}l}
 \hline
 \hline
\multicolumn{2}{c}{$p_T$}  &\multicolumn{2}{c}{${\mit \pi^{-}p}$ @  515~GeV/$c$} \\
\multicolumn{2}{c}{(GeV/$c$)}  &\multicolumn{2}{c}{$(\rm{nb}$/(GeV/$c)^2$)} \\
 \hline
  3.00 & 3.50 & 
69 & 62${}\pm{}$9 \\
  3.50 & 4.00 & 
9.1 & 3.8${}\pm{}$1.1 \\
  4.00 & 4.50 & 
1.76 & 0.31${}\pm{}$0.20 \\
 \hline
\multicolumn{2}{c}{ }  &\multicolumn{2}{c}{$(\rm{pb}$/(GeV/$c)^2$)} \\
 \hline
  4.50 & 5.00 & 
400 & 110${}\pm{}$50 \\
  5.00 & 5.50 & 
121 & 30${}\pm{}$14 \\
  5.50 & 6.00 & 
19 & 17${}\pm{}$2 \\
  6.00 & 7.00 & 
7.8 & 4.7${}\pm{}$0.9 \\
  7.00 & 8.00 & 
0.37 & 0.97${}\pm{}$0.05 \\
 \hline
 \hline
 \end{tabular*}
\label{eta_table_p}
\end{table}

\newpage
\clearpage

\begin{widetext}
\widetext
\begin{table}
\caption{The averaged invariant differential cross section
$\left( \DIFFXS \right)$ per nucleon as a function of rapidity and $p_T$
for the inclusive reaction \piBe~$\rightarrow\eta$X at 515~GeV/$c$.
Units are pb/(GeV/$c$)$^2$.}
\squeezetable
\begin{tabular}{r@{ -- }l
r@{ }l
r@{ }l
r@{ }l}
 \hline
 \hline
\multicolumn{2}{c}{$y_{cm}$}      &\multicolumn{6}{c}{$p_T$ (GeV/$c$)} \\
\multicolumn{2}{c}{ } &\multicolumn{2}{c}{3.00{ -- }4.00 } &\multicolumn{2}{c}{4.00{ -- }4.50 } &\multicolumn{2}{c}{4.50{ -- }5.00 } \\
 \hline         
  $-0.750$ & $-0.625$ & 
\vctr{14000${\,}\pm$} & \vctr{6200${}\pm{}$1700} & {680${\,}\pm$} & {260${}\pm{}$79} & {287${\,}\pm$} & {88${}\pm{}$33} \\
  $-0.625$ & $-0.500$ & 
   &    & {880${\,}\pm$} & {220${}\pm{}$100} & {278${\,}\pm$} & {61${}\pm{}$32} \\
  $-0.500$ & $-0.375$ & 
\vctr{16900${\,}\pm$} & \vctr{5600${}\pm{}$2100} & {1290${\,}\pm$} & {180${}\pm{}$150} & {339${\,}\pm$} & {51${}\pm{}$39} \\
  $-0.375$ & $-0.250$ & 
   &    & {1810${\,}\pm$} & {170${}\pm{}$210} & {456${\,}\pm$} & {45${}\pm{}$53} \\
  $-0.250$ & $-0.125$ & 
\vctr{27100${\,}\pm$} & \vctr{4500${}\pm{}$3300} & {1710${\,}\pm$} & {150${}\pm{}$200} & {580${\,}\pm$} & {53${}\pm{}$67} \\
  $-0.125$ & $0.000$  & 
   &    & {2210${\,}\pm$} & {140${}\pm{}$260} & {556${\,}\pm$} & {47${}\pm{}$64} \\
  $0.000$  & $0.125$  & 
\vctr{27500${\,}\pm$} & \vctr{2300${}\pm{}$3400} & {1750${\,}\pm$} & {130${}\pm{}$200} & {578${\,}\pm$} & {46${}\pm{}$67} \\
  $0.125$  & $0.250$  & 
   &    & {2400${\,}\pm$} & {140${}\pm{}$280} & {650${\,}\pm$} & {55${}\pm{}$75} \\
  $0.250$  & $0.375$  & 
\vctr{23700${\,}\pm$} & \vctr{2400${}\pm{}$2900} & {2360${\,}\pm$} & {150${}\pm{}$270} & {683${\,}\pm$} & {51${}\pm{}$79} \\
  $0.375$  & $0.500$  & 
   &    & {2280${\,}\pm$} & {140${}\pm{}$260} & {688${\,}\pm$} & {51${}\pm{}$80} \\
  $0.500$  & $0.625$  & 
\vctr{18100${\,}\pm$} & \vctr{2100${}\pm{}$2200} & {2030${\,}\pm$} & {120${}\pm{}$240} & {650${\,}\pm$} & {46${}\pm{}$75} \\
  $0.625$  & $0.750$  & 
   &    & {2070${\,}\pm$} & {120${}\pm{}$240} & {595${\,}\pm$} & {43${}\pm{}$69} \\
 \hline         
\multicolumn{2}{c}{ } &\multicolumn{2}{c}{5.00{ -- }5.50 } &\multicolumn{2}{c}{5.50{ -- }6.50 } &\multicolumn{2}{c}{6.50{ -- }8.00 } \\
 \hline         
  $-0.750$ & $-0.625$ & 
{37${\,}\pm$} & {15${}\pm{}$4.3} & {6.8${\,}\pm$} & {4.6${}\pm{}$0.8} & \vctr{0.48${\,}\pm$} & \vctr{0.48${}\pm{}$0.06} \\
  $-0.625$ & $-0.500$ & 
{79${\,}\pm$} & {22${}\pm{}$9.1} & {22.9${\,}\pm$} & {4.8${}\pm{}$2.7} &    &    \\
  $-0.500$ & $-0.375$ & 
{141${\,}\pm$} & {22${}\pm{}$16} & {18.5${\,}\pm$} & {4.3${}\pm{}$2.2} & \vctr{1.31${\,}\pm$} & \vctr{0.48${}\pm{}$0.16} \\
  $-0.375$ & $-0.250$ & 
{125${\,}\pm$} & {17${}\pm{}$15} & {32.2${\,}\pm$} & {5.6${}\pm{}$3.8} &    &    \\
  $-0.250$ & $-0.125$ & 
{157${\,}\pm$} & {17${}\pm{}$18} & {36.5${\,}\pm$} & {4.5${}\pm{}$4.3} & \vctr{3.63${\,}\pm$} & \vctr{0.72${}\pm{}$0.44} \\
  $-0.125$ & $0.000$  & 
{143${\,}\pm$} & {19${}\pm{}$17} & {42.5${\,}\pm$} & {5.5${}\pm{}$5.0} &    &    \\
  $0.000$  & $0.125$  & 
{137${\,}\pm$} & {19${}\pm{}$16} & {33.4${\,}\pm$} & {5.2${}\pm{}$3.9} & \vctr{3.57${\,}\pm$} & \vctr{0.78${}\pm{}$0.43} \\
  $0.125$  & $0.250$  & 
{197${\,}\pm$} & {22${}\pm{}$23} & {45.6${\,}\pm$} & {5.5${}\pm{}$5.4} &    &    \\
  $0.250$  & $0.375$  & 
{225${\,}\pm$} & {21${}\pm{}$26} & {48.0${\,}\pm$} & {5.4${}\pm{}$5.6} & \vctr{4.36${\,}\pm$} & \vctr{0.85${}\pm{}$0.53} \\
  $0.375$  & $0.500$  & 
{218${\,}\pm$} & {22${}\pm{}$25} & {46.0${\,}\pm$} & {6.4${}\pm{}$5.4} &    &    \\
  $0.500$  & $0.625$  & 
{184${\,}\pm$} & {20${}\pm{}$21} & {41.5${\,}\pm$} & {5.8${}\pm{}$4.9} & \vctr{3.86${\,}\pm$} & \vctr{0.77${}\pm{}$0.47} \\
  $0.625$  & $0.750$  & 
{173${\,}\pm$} & {18${}\pm{}$20} & {35.9${\,}\pm$} & {5.1${}\pm{}$4.2} &    &    \\
 \hline
 \hline
 \end{tabular}
\label{eta_table_be_rap}
\end{table}

\begin{table}
\caption{The averaged invariant differential cross section
$\left( \DIFFXS \right)$ as a function of rapidity and $p_T$
for the inclusive reaction \pip~$\rightarrow\eta$X at 515~GeV/$c$.
Units are pb/(GeV/$c$)$^2$.}
\squeezetable
\begin{tabular}{r@{ -- }l
r@{ }l
r@{ }l}
 \hline
 \hline
\multicolumn{2}{c}{$y_{cm}$}      &\multicolumn{4}{c}{$p_T$ (GeV/$c$)} \\
\multicolumn{2}{c}{ } &\multicolumn{2}{c}{4.00{ -- }5.00 } &\multicolumn{2}{c}{5.00{ -- }6.00 } \\
 \hline         
  $-0.75$  & $-0.50$  & 
{550${\,}\pm$} & {550${}\pm{}$64} & \multicolumn{2}{c}{---} \\
  $-0.50$  & $-0.25$  & 
{950${\,}\pm$} & {460${}\pm{}$110} & {74${\,}\pm$} & {42${}\pm{}$8.6} \\
  $-0.25$  & $0.00$   & 
{1170${\,}\pm$} & {360${}\pm{}$140} & {39${\,}\pm$} & {38${}\pm{}$4.5} \\
  $0.00$   & $0.25$   & 
{1840${\,}\pm$} & {430${}\pm{}$210} & {147${\,}\pm$} & {47${}\pm{}$17} \\
  $0.25$   & $0.50$   & 
{1020${\,}\pm$} & {270${}\pm{}$120} & {153${\,}\pm$} & {50${}\pm{}$18} \\
  $0.50$   & $0.75$   & 
{960${\,}\pm$} & {220${}\pm{}$110} & {44${\,}\pm$} & {43${}\pm{}$5.1} \\
 \hline
 \hline
 \end{tabular}

\label{eta_table_p_rap}
\end{table}
\end{widetext}

\vfil
\end{document}